\documentclass[aps,prd,onecolumn,showpacs,showkeys,amsmath,amssymb,article]{revtex4}
\usepackage{epsfig}
\usepackage{amsmath}  
\usepackage{makecell}
\usepackage{graphicx}
\usepackage{array}
\usepackage{subfigure}
\usepackage{bm}
\usepackage{longtable}
\usepackage{tabularx}
\usepackage{booktabs}
\usepackage{appendix}
\usepackage{caption2}
\usepackage{amsfonts}
\usepackage{dcolumn}
\usepackage{siunitx}
\usepackage{tikz-feynman}

\usepackage[utf8]{inputenc}

\graphicspath{{figs/}}
\usepackage{float}
\usepackage{lscape}
\usepackage{txfonts}
\usepackage{overpic}
\usepackage{amssymb}
\usepackage{indentfirst}

\usepackage{feynmf}   
\usepackage{epstopdf}   
\usepackage{slashed}  
\usepackage{multirow}

\usepackage[colorlinks, citecolor=blue,anchorcolor=red,menucolor=red, linkcolor=red,filecolor=red,runcolor=red,urlcolor=blue,frenchlinks=red]{hyperref}
\makeatletter
\newcommand{\figcaption}{\def\@captype{figure}\caption}
\newcommand{\tabcaption}{\def\@captype{table}\caption}

\newcommand{\Rmnum}[1]{\expandafter\@slowromancap\romannumeral #1@}
\def\hlinewd#1{%
  \noalign{\ifnum0=`}\fi\hrule \@height #1 \futurelet
   \reserved@a\@xhline}
\makeatother

\begin{document}
\title{Doubly charmed pentaquark states in QCD sum rules}

\author{Feng-Bo Duan$^1$}
\author{Qi-Nan Wang$^1$}
\author{Zi-Yan Yang$^{3,\, 4}$}
\author{Xu-Liang Chen$^1$}
\author{Wei Chen$^{1,\, 2}$}
\email{chenwei29@mail.sysu.edu.cn}
\affiliation{$^1$School of Physics, Sun Yat-sen University, Guangzhou 510275, China \\ 
$^2$Southern Center for Nuclear-Science Theory (SCNT), Institute of Modern Physics, 
Chinese Academy of Sciences, Huizhou 516000, Guangdong Province, China\\
$^3$Key Laboratory of Atomic and Subatomic Structure and Quantum Control (MOE), Guangdong Basic Research Center of Excellence for Structure and Fundamental Interactions of Matter, Institute of Quantum Matter, South China Normal University, Guangzhou 510006, China\\
$^4$Guangdong-Hong Kong Joint Laboratory of Quantum Matter, Guangdong Provincial Key Laboratory of Nuclear Science, Southern Nuclear Science Computing Center, South China Normal University, Guangzhou 510006, China}

\begin{abstract}
We have studied the mass spectra of doubly charmed pentaquark states in the $\Lambda  _{c}^{(*)}D^{(*)}$ and $\Sigma _{c}^{(*)}D^{(*)}$ channels with $J^P=1/2^\pm$, $3/2^\pm$ and $5/2^\pm$ within the framework of QCD sum rules. We use the parity projected sum rules to separate the contributions of negative and positive parities from the two-point correlations induced by the pentaquark interpolating currents. Our results show that the bound states of $P_{cc}$ pentaquarks may exist in the $\Lambda _cD\, (\frac{1}{2}^-)$, $\Sigma _cD\, (\frac{1}{2}^-)$, $\Sigma _cD^*\, (\frac{3}{2}^-)$, $\Lambda _c^*D\, (\frac{3}{2}^-)$, $\Lambda _c^*D^*\, (\frac{5}{2}^-)$ channels with negative-parity and $\Sigma _cD\, (\frac{1}{2}^+)$, $\Sigma _cD^\ast\, (\frac{3}{2}^+)$, $\Sigma _c^\ast D\, (\frac{3}{2}^+)$ channels with positive-parity, since their masses are predicted to be lower than the corresponding meson-baryon thresholds. However, they are still allowed to decay into the $\Xi_{cc}^{(\ast)}\pi$ final states via strong interaction. The triply charged $P_{cc}^{+++}(ccuu\bar d)$ and neutral $P_{cc}^{0}(ccdd\bar u)$ in the isospin quartet would definitely be pentaquark states due to their exotic charges. We suggest searching for these characteristic doubly charmed pentaquark signals in the $P_{cc}^{+++}\to\Xi_{cc}^{(\ast) ++}\pi^+/\rho^+$, $\Sigma_c^{(\ast)++}D^{(\ast)+}$ and $P_{cc}^{0}\to\Xi_{cc}^{(\ast) +}\pi^-/\rho^-$, $\Sigma_c^{(\ast)0}D^{(\ast)0}$ decays in the near future.
\end{abstract}
\pacs{12.39.Mk, 12.38.Lg, 14.20.Lq}
\keywords{Doubly charmed states, Pentaquarks, QCD sum rules}
\maketitle

\section{Introduction}\label{Sec:Intro}
Since the discovery of X(3872) in 2003, there have been reported numerous new hadron states, including the XYZ states, hidden-charm pentaquark states, doubly charmed tetraquark $T_{cc}$ state, fully-charm tetraquark states, doubly charged $T_{c\bar s}(2900)^{++}$ state and so on~\cite{Chen:2016qju,Guo:2017jvc,Liu:2019zoy,Brambilla:2019esw,Chen:2022asf,Chen:2020uif}. They are  good candidates of exotic hadron states. It is very important to study their various properties for identifying and understanding the new hadron spectroscopy and nonperturbative QCD. 

In 2015, the LHCb Collaboration observed two hidden-charm pentaquark states $P_{c}(4380)$ and $P_{c}(4450)$ in the $J/\psi p$ invariant mass spectrum in the $\Lambda _b^0\to J/\psi K^- p$ process~\cite{Aaij2015}. Combined with LHCb's Run 2 data in 2019, they  discovered a new narrow pentaquark $P_{c}(4312)$ in the same process and final states~\cite{Aaij2019}. They further claimed that $P_{c}(4450)$ was consisted of two narrow pentaquark structures $P_{c}(4440)$ and $P_{c}(4457)$ with the statistical significance of $5.4\sigma$~\cite{Aaij2019}. In the $B_{s}^0\to J/\psi \bar{p}p$ decays, an evidence for a new structure $P_{c}(4337)$ was found in both the $J/\psi p$ and $J/\psi \bar{p}$ systems~\cite{Aaij2022}. The story was ongoing with the observation of $P_{cs}(4459)$, which is the candidate of hidden-charm pentaquark state with strangeness reported in the $J/\psi \Lambda$ invariant mass distribution in the $\Xi_{b}^-\to J/\psi K^- \Lambda$ decays~\cite{Aaij2021}. Very recently, LHCb also reported a very narrow resonance $P_{cs}(4338)$ in the $B^-\to J/\psi \Lambda \bar{p}$ process, with the preferred spin-parity quantum numbers $J^P=1/2^-$ at high confidence level~\cite{LHCb:2022ogu}. 

The observations of these $P_c$ and $P_{cs}$ structures have attracted extensive theoretical studies in the hidden-charm pentaquark states, including their inner structures, quantum numbers, mass spectra, decay and production properties, etc. On the other hand, one may wonder if there exist pentaquark states other than in the hidden-charm channels, such as the doubly charmed pentaquarks. Such expectations were directly inspired by the observations of the doubly charmed baryons and doubly charmed tetraquark state. In 2017, the LHCb Collaboration discovered the doubly charmed baryon $\Xi _{cc}^{++}$ in the $\Lambda _{c}^+ K^-\pi ^+\pi ^+$ mass spectrum~\cite{Aaij2017} and confirmed it later in the $\Xi _{cc}^{++}\to\Xi _{c}^{+}\pi^+$ decay mode~\cite{LHCb:2018pcs}, shedding light on the long-standing puzzle in the $\Xi _{cc}$ system after the previous observations of the $\Xi _{cc}^{+}$ baryon by SELEX in the $\Lambda _{c}^+ K^-\pi ^+$ final states~\cite{SELEX:2002wqn,SELEX:2004lln}. Recently in 2022, LHCb reported the observation of the doubly charmed tetraquark state $T_{cc}^+$ in the mass spectrum of $D^0D^0\pi^+$ just below the $D^{\ast +}D^0$ threshold, with very narrow decay width and the quantum numbers $I(J^P)=0(1^+)$~\cite{Aaij2022a,Aaij2022b}. 

To date, there are some theoretical investigations on the existence of doubly charmed pentaquark states~\cite{Chen2017,Liu2020,Chen2021b,Chen2021a,Dong2021a,Shen2023,Liu:2023clr,Yang:2020twg,Shimizu:2017xrg,Guo:2017vcf,Zhou:2018bkn,Wang:2018lhz,Ozdem:2022vip,Xing:2021yid}. 
In Ref.~\cite{Chen2017},the authors explored the intermediate- and short-range forces from $\sigma/\omega$ exchange in the one-boson-exchange (OBE) model, and suggested that the interaction in the S-wave $\Lambda_c D$ system be attractive to form a molecular state. In Refs.~\cite{Liu2020,Chen2021b}, the doubly charmed $D^{(\ast)}\Sigma_c^{(\ast)}$ pentaquark systems were investigated in the OBE models and the attractive interactions were found for all channels with isospin $I=1/2$ so that the molecular bound states can be formed. Such results were also supported by the systematic studies within the framework of chiral effective field theory~\cite{Chen2021a}, Bethe-Salpeter approach~\cite{Dong2021a}, and the unitarized coupled-channel approach~\cite{Shen2023}. Using the resonating group method in the QDCSM framework~\cite{Liu:2023clr}, the doubly charmed $\Lambda_c/\Sigma_c^{(*)}D^{(*)}$, $\Xi_{cc}^{(*)}\pi/\eta/\rho/\omega$ pentaquark systems were systematically investigated, in which a possible bound state $\Xi_{cc}^\ast\rho$ with $I(J^P)=3/2(5/2^{-})$ and five resonance states with various quantum numbers were obtained. Another shallow bound state $\Xi_{cc}^\ast\pi$ with $I(J^P)=1/2(3/2^{-})$ and several possible narrow $\Sigma_c^{(*)} D^{(*)}$ resonances were predicted within the chiral quark model~\cite{Yang:2020twg}. In Ref.~\cite{Shimizu:2017xrg}, the coupled system of $\Lambda_c D^{(*)}-\Sigma_c^{(*)} D^{(*)}$ in $I(J^P)=1/2(3/2^{-})$ channel was studied and a doubly charmed pentaquark state was found as a hadronic molecule. Within the chiral effective field theory, the authors in Ref.~\cite{Guo:2017vcf} studied the S-wave scatterings of doubly charmed baryons and light pseudoscalar mesons, and they found one bound state below the $\Xi_{cc}\bar K$ threshold with $(S,\, I)=(-1,\, 0)$ and two resonant structures around the $\Xi_{cc}\pi$, $\Xi_{cc}\eta$, $\Omega_{cc} K$ thresholds with $(S,\, I)=(0,\, 1/2)$. Besides, the doubly charmed pentaquarks have been also investigated in the compact configurations of diquark-diquark-antiquark model~\cite{Zhou:2018bkn,Wang:2018lhz,Ozdem:2022vip} and triquark-diquark model~\cite{Xing:2021yid}, including their masses, lifetimes, magnetic moments and so on. In this work, we shall systematically study the possible doubly charmed pentaquark states in the $\Lambda  _{c}^{(*)}D^{(*)}$ and $\varSigma _{c}^{(*)}D^{(*)}$ channels with $J^P=1/2^\pm$, $3/2^\pm$ and $5/2^\pm$ using the QCD sum rule method. 


This paper is organized as follows. In Sec. II, we construct the interpolating currents for $\Lambda  _{c}^{(*)}D^{(*)}$ and $\varSigma_{c}^{(*)}D^{(*)}$ meson-baryon $P_{cc}$ states. In Sec. III, we establish the parity projected sum rules for these pentaquark systems and calculate the correlation functions and spectral densities for all interpolating currents.
We extract the masses and coupling constants for these doubly charmed pentaquark states by performing the QCD sum rule analyses in Sec. IV. The last section is a brief summary.

\begin{table}[t!]
  \caption{Masses and spin-parity quantum numbers for charmed baryons and charmed mesons.}\label{mesonbaryonTab}
  \renewcommand\arraystretch{1.8} 
  \setlength{\tabcolsep}{0.5 em}{ 
  \begin{tabular}{c c c c c c c c c c c}
    \hline
     \hline
State & $\Lambda _c$ & $\Lambda _c(2595)$ & $\Lambda _c(2860)/\Lambda _c^*$ & $\Lambda _c(2625)/\Lambda _c^*$ & $\Sigma _c$ & $\Sigma _c$  & $\Sigma _c(2520)/\Sigma _c^*$ & $\Sigma _c(2800)/\Sigma _c^*$ & $D$ & $D^*$   \\ \hline
$J^P$ & $\frac{1}{2}^+$ & $\frac{1}{2}^-$ & $\frac{3}{2}^+$ & $\frac{3}{2}^-$ & $\frac{1}{2}^+$ & $\frac{1}{2}^-$  & $\frac{3}{2}^+$ & $\frac{3}{2}^-$ & $0^-$ &$1^-$  \\
Mass ($\mathrm{MeV}$)   & 2286 & 2592 & 2860  & 2625 & 2455 & 2770 & 2520 & 2805 & 1869 & 2007 \vspace{1ex} \\
   \hline
     \hline
  \end{tabular}
  }
  \end{table}

\section{QCD  sum rules for doubly charmed pentaquarks}\label{Sec:Current}
In this section, we construct the interpolating currents for the doubly charmed pentaquark states in the $[qcq][\bar qc]$ baryon-meson molecular picture, where $q=(u,\, d)$ is a light quark and $c$ is the charm quark. To investigate the hadron molecules in the $\Lambda  _{c}^{(*)}D^{(*)}$ and $\Sigma _{c}^{(*)}D^{(*)}$ channels, we firstly collect the masses and spin-parity for the charmed baryons $\Lambda  _{c}^{(*)}, \Sigma _{c}^{(*)}$ and charmed mesons $D^{(*)}$ in Table~\ref{mesonbaryonTab} from PDG~\cite{ParticleDataGroup:2022pth}. For the negative-parity $\Sigma_c/\Sigma_c^{\ast}$ with $\frac{1}{2}^-/\frac{3}{2}^-$, we use the theoretical predicted masses in the relativized potential quark model~\cite{Capstick:1986ter}. Replacing the $s$ quark to $c$ quark in the well-known Ioffe currents~\cite{Ioffe1981}, we construct the interpolating currents for the $\Lambda_c^{(\ast)}$ and $\Sigma_c^{(\ast)}$ charmed baryons 
\begin{equation}\label{baryoncurrents}
    \begin{split}
      j^{\Lambda_{c}}&=\varepsilon ^{abc}\left[\left(u_{a}^{T}\mathcal{C} \gamma _{\nu }c_{b} \right)\gamma _{5}\gamma ^{\nu }d_{c}-\left(d_{a}^{T}\mathcal{C} \gamma _{\nu }c_{b} \right)\gamma _{5}\gamma ^{\nu }u_{c}\right]\, ,\\
      j_{\nu}^{\Lambda_{c}^*}&=\varepsilon ^{abc}\left[\left(u_{a}^{T}\mathcal{C} \gamma _{\nu }c_{b} \right)d_{c}-\left(u_{a}^{T}\mathcal{C} \gamma _{\nu }d_{b} \right)c_{c}\right] \, ,\\
      j^{\varSigma_{c}}&=\varepsilon ^{abc}\left[\left(u_{a}^{T}\mathcal{C} \gamma _{\nu }c_{b} \right)\gamma _{5}\gamma ^{\nu }d_{c}+\left(d_{a}^{T}\mathcal{C} \gamma _{\nu }c_{b} \right)\gamma _{5}\gamma ^{\nu }u_{c}\right] \, ,\\
      j_{\nu }^{\varSigma_{c}^*}&=\varepsilon ^{abc}\left[2\left(u_{a}^{T}\mathcal{C} \gamma _{\nu }c_{b} \right)u_{c}+\left(u_{a}^{T}\mathcal{C} \gamma _{\nu }u_{b} \right)c_{c}\right]\, ,\\  
          \end{split}
\end{equation}
where the superscripts and subscripts $a, b, c$ are the color indices, the superscript $T$ is the transposition operator, and $\mathcal{C} $ is the charge conjugation matrix. The interpolating current $j_{\nu}^{\Lambda_{c}^*}$ in Eq.~\eqref{baryoncurrents} for $\Lambda_{c}^*$ baryon can be obtained  using the SU(2) symmetry relations. These currents can couple to both the positive and negative parity charmed baryons corresponding to S-wave and P-wave states respectively. 
For the charmed mesons, only the $S$-wave $D^{(\ast)}$ fields are considered 
\begin{equation}\label{mesoncurrents}
    \begin{split}
      j^{D}=\bar{d}_{d}i\gamma _{5}c_{d}\, , \,\,
      j_\mu^{D^\ast}=\bar{d}_{d}\gamma _{\mu}c_{d}\, .\\  
          \end{split}
\end{equation}

Using the above charmed baryon and meson operators, we construct the interpolating currents for the doubly charmed $\Lambda _{c}^{(*)}D^{(*)}$ and $\Sigma _{c}^{(*)}D^{(*)}$ molecular pentaquarks as the following
\begin{equation}\label{currents1}
    \begin{split}
      &J^{\Lambda  _{c}D}=\varepsilon ^{abc}\left[\left(u_{a}^{T}\mathcal{C} \gamma _{\mu }c_{b} \right)\gamma _{5}\gamma ^{\mu }d_{c}-\left(d_{a}^{T}\mathcal{C} \gamma _{\mu }c_{b} \right)\gamma _{5}\gamma ^{\mu }u_{c}\right] \left[\bar{d}_{d}i\gamma _{5}c_{d}\right] \, ,\\
      &J_{\mu }^{\Lambda  _{c}D^*}=\varepsilon ^{abc}\left[\left(u_{a}^{T}\mathcal{C} \gamma _{\nu }c_{b} \right)\gamma _{5}\gamma ^{\nu }d_{c}-\left(d_{a}^{T}\mathcal{C} \gamma _{\nu }c_{b} \right)\gamma _{5}\gamma ^{\nu }u_{c}\right] \left[\bar{d}_{d}\gamma_{\mu }c_{d}\right] \, ,\\
      &J_{\mu }^{\Lambda  _{c}^*D}=\varepsilon ^{abc}\left[\left(u_{a}^{T}\mathcal{C} \gamma _{\mu }c_{b} \right)d_{c}-\left(u_{a}^{T}\mathcal{C} \gamma _{\mu }d_{b} \right)c_{c}\right] \left[\bar{d}_{d}i\gamma _{5}c_{d}\right] \, ,\\
      &J_{\mu \nu }^{\Lambda  _{c}^*D^*}=\varepsilon ^{abc}\left[\left(u_{a}^{T}\mathcal{C} \gamma _{\nu }c_{b} \right)d_{c}-\left(u_{a}^{T}\mathcal{C} \gamma _{\nu }d_{b} \right)c_{c}\right] \left[\bar{d}_{d}\gamma_{\mu }c_{d}\right]+\left(\mu \leftrightarrow \nu \right) \, ,\\
      &J^{\varSigma  _{c}D}=\varepsilon ^{abc}\left[\left(u_{a}^{T}\mathcal{C} \gamma _{\mu }c_{b} \right)\gamma _{5}\gamma ^{\mu }d_{c}+\left(d_{a}^{T}\mathcal{C} \gamma _{\mu }c_{b} \right)\gamma _{5}\gamma ^{\mu }u_{c}\right] \left[\bar{d}_{d}i\gamma _{5}c_{d}\right] \, ,\\
      &J_{\mu }^{\varSigma  _{c}D^*}=\varepsilon ^{abc}\left[\left(u_{a}^{T}\mathcal{C} \gamma _{\nu }c_{b} \right)\gamma _{5}\gamma ^{\nu }d_{c}+\left(d_{a}^{T}\mathcal{C} \gamma _{\nu }c_{b} \right)\gamma _{5}\gamma ^{\nu }u_{c}\right] \left[\bar{d}_{d}\gamma_ {\mu }c_{d}\right] \, ,\\
      &J_{\mu }^{\varSigma  _{c}^*D}=\varepsilon ^{abc}\left[2\left(u_{a}^{T}\mathcal{C} \gamma _{\mu }c_{b} \right)u_{c}+\left(u_{a}^{T}\mathcal{C} \gamma _{\mu }u_{b} \right)c_{c}\right] \left[\bar{d}_{d}i\gamma _{5}c_{d}\right] \, ,\\
      &J_{\mu \nu }^{\varSigma  _{c}^*D^*}=\varepsilon ^{abc}\left[2\left(u_{a}^{T}\mathcal{C} \gamma _{\nu }c_{b} \right)u_{c}+\left(u_{a}^{T}\mathcal{C} \gamma _{\nu }u_{b} \right)c_{c}\right] \left[\bar{d}_{d}\gamma_ {\mu }c_{d}\right]+\left(\mu \leftrightarrow \nu \right) \, .\\  
          \end{split}
\end{equation}
These pentaquark interpolating currents $J$, $J_{\mu}$, and $J_{\mu\nu}$ carry the spin-parity quantum numbers $J^P=1/2^-$, $3/2^-$, and $5/2^-$, respectively. Besides, the current $J_{\mu\nu}$ should also contain the $J=1/2$ and $3/2$ components while $J_{\mu}$ contains the $J=1/2$ component. 
Similar to the baryonic current, a pentaquark operator with negative-parity can couple to both the negative-parity and positive-parity pentaquark states via different coupling relations~\cite{Chung1982,Bagan1993,Jido1996,Ohtani:2012ps}
\begin{equation}\label{couple1}
    \begin{split}
      &\langle 0|J_-|X_{1/2^-}\rangle=f_{X}^-u(p)\, ,\\
         &\langle 0|J_-|X_{1/2^+}\rangle=f_{X}^+\gamma_5u(p)\, ,\\
          \end{split}
\end{equation}
in which $u(p)$ is the Dirac spinor and $f_X^\mp$ is the coupling constant. 
The spin-parity for the current $J_-$ is $J^P=1/2^-$ under the non-$\gamma_5$ coupling. It can also couple to the positive-parity pentaquark via the above $\gamma_5$ coupling relation. These relations also suggest that the current $J_{+}\equiv i\gamma _{5}J_{-}$ with opposite parity can couple to the same pentaquark states with $J_-$, so that leading to the same sum rule results. For the vector and tensor currents, the non-$\gamma_5$ couplings to the hadron states are 
\begin{equation}\label{couple2}
    \begin{split}
      &\langle 0|J_{\mu }|X_{\frac{3}{2}}\rangle=f_{X}u_{\mu }(p)\, ,\\
      &\langle 0|J_{\mu \nu }|X_{\frac{5}{2}}\rangle=f_{X}u_{\mu \nu }(p)\, ,\\
          \end{split}
\end{equation}
where $u_{\mu }(p)$ and $u_{\mu \nu }(p)$ are the Rarita-Schwinger vector and polarization tensor, respectively.

In QCD sum rules, the two-point correlation functions $\Pi (p)$, $\Pi_{\mu \nu } (p)$ and $\Pi_{\mu \nu \alpha \beta } (p)$ can be written as~\cite{Shifman1979,Reinders1985,Colangelo:2000dp},
\begin{equation}\label{correlation}
    \begin{split}
    \Pi(p^2)&=i\int  d^4x e^{ip\cdot x}\langle 0|T\left[J(x) \bar{J}(0)\right]|0\rangle\\
         &=\Pi^{1/2}(p^2)\,,
    \end{split}
    \end{equation}
\begin{equation}\label{correlation1}
    \begin{split}
    \Pi_{\mu\nu }(p^2)&=i\int d^4x e^{ip\cdot x}\langle 0|T\left[J_{\mu}(x) \bar{J}_{\nu }(0)\right]|0\rangle\\
         &=\left(\frac{p_{\mu}p_{\nu}}{p^2 }-g_{\mu\nu }\right) \Pi^{3/2}(p^2)+\cdots\, ,
    \end{split}
    \end{equation}
\begin{equation}\label{correlation2}
    \begin{split}
    \Pi_{\mu\nu\alpha \beta }(p^2)&=i\int d^4x e^{ip\cdot x}\langle 0|T\left[J_{\mu\nu}(x) \bar{J}_{\alpha \beta }(0)\right]|0\rangle\\
    &=\left(g_{\mu\alpha } g_{\nu\beta }+g_{\mu\beta }g_{\nu\alpha }\right) \Pi^{5/2}(p^2)+\cdots \, ,
    \end{split}
    \end{equation}
in which $\hat p=\gamma_\mu p^\mu$. The $\Pi^{1/2}(p^2)$, $\Pi^{3/2}(p^2)$ and $\Pi^{5/2}(p^2)$ are the invariant functions for the hadron states with spin-1/2, 3/2 and 5/2 respectively. 
In Eqs.~\eqref{correlation1}-\eqref{correlation2}, there are also some other contributions from the spin-1/2 and 3/2 components, as mentioned above. 
    In this article, we choose the tensor structures 1, $g_{\mu \nu}$ and
$g_{\mu\alpha } g_{\nu\beta }+g_{\mu\beta }g_{\nu\alpha }$ for the correlation functions $\Pi(p^2)$, $\Pi_{\mu\nu }(p^2)$
and $\Pi_{\mu\nu\alpha \beta }(p^2)$ respectively to study the $J^P=1/2^\pm $, $3/2^\pm $ and $5/2^\pm$ doubly
heavy pentaquark states. 

We calculate the invariant function $\Pi ^{1/2}(p^2)$, $\Pi ^{3/2}(p^2)$ and $\Pi ^{5/2}(p^2)$ 
at both hadronic and quark-gluonic levels. At the hadronic level, we use the dispersion relation to write the invariant function as:
\begin{equation}\label{correlationhad}
    \begin{split}
    \Pi(p^2)&=\frac{1}{\pi } \int^{\infty }_{4m_c^2} \frac{\text{Im}\Pi (s)}{s-p^2-i\varepsilon  }ds\,,
    \end{split}
    \end{equation}
   where $4m_c^2$ is the physical threshold. The imaginary part of the invariant function $\text{Im}\Pi (s)$ is related to the spectral function $\rho(s)\equiv\text{Im}\Pi (s)/\pi$, which is estimated at the phenomenological hadronic level by
    inserting intermediate pentaquark states $\sum _{n}\vert n \rangle\langle n \vert $~\cite{Shifman1979,Reinders1985,Colangelo:2000dp}
\begin{equation}\label{spectral}
        \begin{split}
          \rho _{phen}(s) \equiv \frac{\text{Im}\Pi (s)}{\pi   }&=\sum_{n }\delta (s-M_{n}^2) \langle 0|J|n \rangle \langle n|\bar{J}|0 \rangle  \\
          &={f_{X}^{-}}^2(\hat p+M^-_X)\delta (s-{M_{X}^{-}}^2)+{f_{X}^{+}}^2(\hat p-M^+_X)\delta (s-{M_{X}^{+}}^2)+\cdots\, .\\
        \end{split}
    \end{equation}
In the last step, we use the ``narrow resonance'' approximation for the ground state, while ``$\cdots$'' contains the contributions from the continuum and excited states. 

It is obvious that the invariant functions in Eqs.~\eqref{correlation}-\eqref{correlation2} contain information of both negative-parity and positive-parity pentaquarks
\begin{equation} \label{correlationpn}
\Pi(p^2)={f_{X}^{-}}^2\frac{\hat p+M_X^-}{{M_{X}^{-}}^2-p^2}+f_{X}^{+2}\frac{\hat p-M_X^+}{{M_{X}^{+}}^2-p^2}+\cdots,\, 
\end{equation}
in which the first and second terms at the right side of Eq.~\eqref{correlationpn} correspond to the negative and positive-parity lowest lying resonances, respectively. It is interesting to see that the only difference of these two parts is the relative sign between $\hat{p}$ and $M_X^\pm$. 
One can isolate the pole terms of the lowest states and obtain the hadronic spectral densities in the rest frame $\overrightarrow{p}=0$~\cite{Jido1996}

\begin{equation}\label{spectralhadron}
  \begin{split}
    \frac{\text{Im}\Pi (p_0)}{\pi   }=&{f_{X}^{-}}^2(\gamma _{0}p_0+M^-_X)\delta (s-{M_{X}^{-}}^2)+f_{X}^{+2}(\gamma _{0}p_0-M^+_X)\delta (s-{M_{X}^{+}}^2)  \\
    =& \gamma _{0}p_0  \rho _{H}^1(s) + \rho _{H}^0(s), \\
    \end{split}
\end{equation}
where

\begin{equation}\label{spectral2}
  \begin{split}
    p_0\rho _{H}^1(s)=& {f_{X}^{-}}^2M^-_X\delta (s-{M_{X}^{-}}^2)+f_{X}^{+2}M^+_X\delta (s-{M_{X}^{+}}^2)\, ,  \\
    \rho _{H}^0(s)=& {f_{X}^{-}}^2M^-_X\delta (s-{M_{X}^{-}}^2)-f_{X}^{+2}M^+_X\delta (s-{M_{X}^{+}}^2)\,,  \\
    \end{split}
\end{equation}
so that the $p_0\rho _{H}^1(s)+\rho _{H}^0(s)$ and $p_0\rho_{H}^1(s)-\rho _{H}^0(s)$ contain only the contribution
from the negative-parity and positive-parity
pentaquark states, respectively.
Now we can obtain two QCD sum rules at the hadronic side
\begin{equation}\label{srhadron1}
  \begin{split}
    \int _{4m_{c}^2}^{s_{0}} \left[\sqrt{s} \rho _{j,H}^1(s)+\rho _{j,H}^0(s) \right] \text{exp}\left(-\frac{s}{M_B^2}\right)ds=2M^-_Xf_{j,X}^{-2}\text{exp}\left(-\frac{{M_{X}^{-}}^2}{M_B^2}\right) \, ,
     \end{split}
\end{equation}
\begin{equation}\label{srhadron2}
  \begin{split}
    \int _{4m_{c}^2}^{s_{0}} \left[\sqrt{s} \rho _{j,H}^1(s)-\rho _{j,H}^0(s) \right] \text{exp}\left(-\frac{s}{M_B^2}\right)ds=2M^+_Xf_{j,X}^{+2}\text{exp}\left(-\frac{{M_{X}^{+}}^2}{M_B^2}\right) \, ,
     \end{split}
\end{equation}
where $j=1/2,\, 3/2,\, 5/2$ is the spin of the pentaquark state.
The parameters $s_0$ and $M_B$ are the continuum threshold and Borel mass, respectively. Eq.~\eqref{srhadron1} is the sum rule for the negative-parity pentaquark while Eq.~\eqref{srhadron2} for the positive-parity one. 

At the quark-gluonic level, the two-point correlation function can be evaluated via the operator production expansion (OPE) method. It is usually expressed as a function of various QCD parameters, such as QCD condensates, quark masses and the
strong coupling constant $\alpha_{s}$. To calculate the Wilson coefficients, we adopt the coordinate space expression for the light quark propagator while the momentum space expression for the heavy quark propagator
    \begin{eqnarray}
\nonumber i S_{q}^{ab}(x)&=&\frac{i\delta^{ab}}{2\pi^2x^4}\hat{x}-\frac{\delta^{ab}}{12}\langle\bar{q}q\rangle+\frac{i}{32\pi^2}\frac{\lambda^n_{ab}}{2}g_sG^n_{\mu\nu}\frac{1}{x^2}(\sigma^{\mu\nu}\hat{x}+\hat{x}\sigma^{\mu\nu})\\
        & &+\frac{\delta^{ab}x^2}{192}\langle\bar{q}g_s\sigma\cdot Gq\rangle-\frac{m\delta^{ab}}{4\pi^2x^2}+\frac{i\delta^{ab}m\langle\bar{q}q\rangle}{48}\hat{x}-\frac{im\langle\bar{q}g_s\sigma\cdot Gq\rangle\delta^{ab}x^2\hat{x}}{1152}\, ,\\
 i S_{Q}^{a b}(p)&=&\frac{i \delta^{a b}}{\hat{p}-m_{Q}}
        +\frac{i}{4} g_{s} \frac{\lambda_{a b}^{n}}{2} G_{\mu \nu}^{n} \frac{\sigma^{\mu \nu}\left(\hat{p}+m_{Q}\right)+\left(\hat{p}+m_{Q}\right) \sigma^{\mu \nu}}{(p^{2}-m_{Q}^{2})^2}
        +\frac{i \delta^{a b}}{12}\left\langle g_{s}^{2} G G\right\rangle m_{Q} \frac{p^{2}+m_{Q} \hat{p}}{(p^{2}-m_{Q}^{2})^{4}}\, ,
       \end{eqnarray}
where $\lambda ^n$ is the Gell-Mann matrix~\cite{Pascual1984}. 
In this work, we calculate correlation functions and spectral densities up to  dimension 10 condensates at the leading order of $\alpha _{s}$. The OPE series contain the perturbative term, the quark condensates $\langle \bar{q}q \rangle$ , the gluon condensate $\langle g_{s}^2GG \rangle$, the
quark-gluon mixed condensates $\langle g_{s} \bar{q} \sigma Gq \rangle$, and the higher dimensional condensates $\langle \bar{q}q \rangle^2$, $\langle \bar{q}q \rangle \langle g_{s} \bar{q} \sigma Gq \rangle$ and $\langle g_{s} \bar{q} \sigma Gq \rangle^2$. The results for the spectral densities $\rho _{D}^{1}(s)$ and $\rho _{D}^{0}(s)$ are shown in the appendix~\ref{appendix}. 
We ignore the chiral suppressed terms with the up and down quark masses, and adopt the factorization assumption of vacuum saturation for higher dimensional condensates.

By equating the invariant functions at two sides, one can obtain the QCD sum rules for the pentaquark mass 
%
\begin{equation}\label{massQCD}
  \begin{split}
     M_{j, \pm }^2=\frac{    \int _{4m_{c}^2}^{s_{0}} \left[\sqrt{s} \rho _{j,QCD}^1(s)\mp \rho _{j,QCD}^0(s) \right] \text{exp}\left(-\frac{s}{M_B^2}\right)sds}
     {\int _{4m_{c}^2}^{s_{0}} \left[\sqrt{s} \rho _{j,QCD}^1(s)\mp \rho _{j,QCD}^0(s) \right] \text{exp}\left(-\frac{s}{M_B^2}\right)ds}\, ,
        \end{split}
\end{equation}
in which $M_{j, +}$ is the mass for the positive-parity pentaquark, $M_{j, -}$ for the negative-parity pentaquark.
  
Considering the isospin quantum number, the $\Lambda _{c}^{(*)}D^{(*)}$ pentaquarks form isospin doublet with $I=1/2$ while the $\Sigma _{c}^{(*)}D^{(*)}$ states can be doublet or quartet with $I=3/2$. We denote the isospin doublet and quartet for the doubly charmed pentaquarks as $(P_{cc}^{++}, \, P_{cc}^{+})$ and $(P_{cc}^{+++}, \, P_{cc}^{++}, \, P_{cc}^{+}, \, P_{cc}^{0})$, respectively. Since we don't differentiate the up and down quarks in our calculations, the predicted pentaquark masses with the same spin-parity quantum numbers should be degenerate. One notes that the states in the isospin quartet are definite exotic pentaquark states since the ordinary doubly charmed baryons ($ccu/ccd$) belong to an isospin doublet. Especially for the triply charged $P_{cc}^{+++}(ccuu\bar d)$ and neutral $P_{cc}^{0}(ccdd\bar u)$ states, they shall be the characteristic signals for searching for these doubly charmed pentaquarks in the future.

\section{Numerical results and discussions}\label{Sec:Numerical}

In this section, we perform the QCD sum rule numerical analyses to predict the masses for the doubly charmed pentaquark states.
We use the following values for various input parameters~\cite{Narison1988,Jamin2002,Jamin1999,Ioffe1981,Chung1984,Dosch1989,Khodjamirian2011,Tanabashi2018,Francis2019,Narison2018}:
\begin{equation}\label{inputparameter}
    \begin{split}
      &m_c(m_c)=1.27^{+0.03}_{-0.04}\;\mathrm{GeV},\\
      &\langle \bar{q}q\rangle=-(0.24\pm0.01)^3\;\mathrm{GeV}^3,\\
      &\langle \bar{q}g_s\sigma\cdot Gq\rangle=-M_0^2\langle \bar{q}q\rangle,\\
      &M_0^2=(0.8\pm0.2)\;\mathrm{GeV}^2,\\
      &\langle g_s^2GG\rangle=(0.48\pm0.14)\;\mathrm{GeV}^4,\\ 
    \end{split}
    \end{equation}  
in which the running mass in the $\overline{MS}$ scheme is used for the charm quark. 
 \begin{figure}[t!]
  \centering
  \includegraphics[width=8cm]{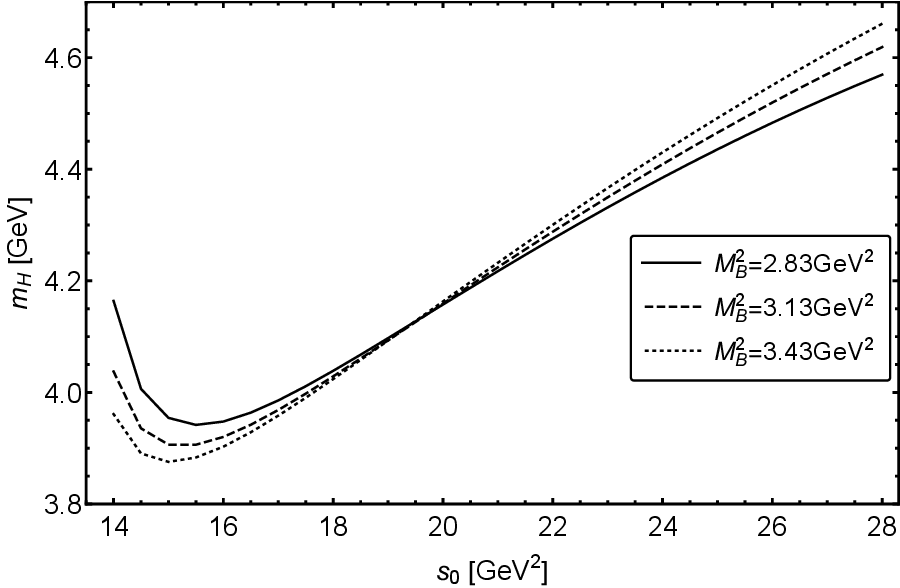}\quad
  \includegraphics[width=8cm]{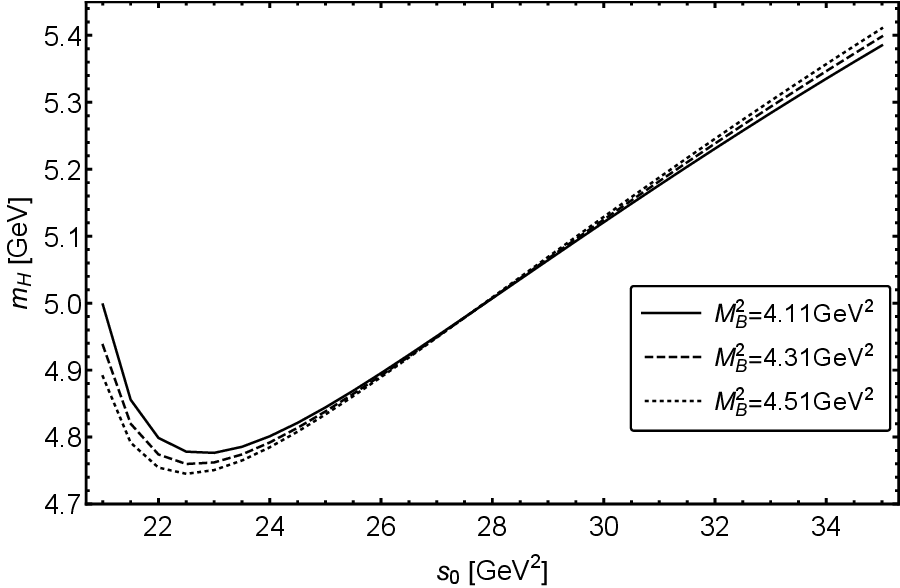}\\
  \caption{Variation of hadron mass to $s_0$ for the current $J^{\Lambda _cD}$ with
  $J^P=\frac{1}{2}^-$ (left) and $J^P=\frac{1}{2}^+$ (right).
    }
  \label{fig:lanmdc-D-mass-s0}
  \end{figure}
     \begin{figure}[t!]
  \centering
  \includegraphics[width=8cm]{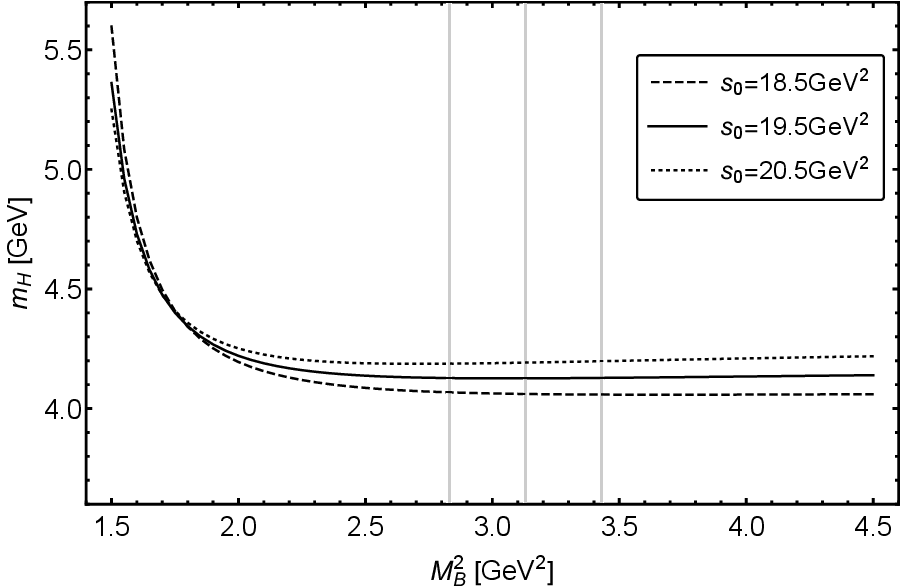}\quad
  \includegraphics[width=8cm]{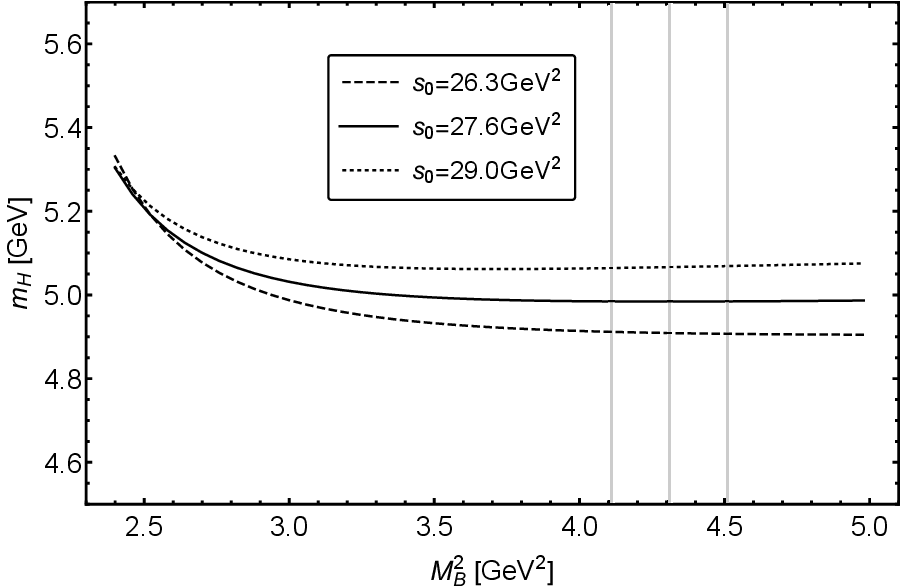}\\
  \caption{Mass curves for the current $J^{\varLambda _cD}$ with $J^P=\frac{1}{2}^-$ (left) and $J^P=\frac{1}{2}^+$ (right).}
  \label{fig:BorelplatformLambdacD}
  \end{figure}

To constrain the working regions of the continuum threshold $s_0$ and Borel mass $M_B$, we consider the behaviors of the OPE convergence (CVG) and pole contribution (PC) as the following
\begin{equation}\label{CVG}
      \begin{split}
    \text{CVG}_{\pm } \equiv  \left\lvert \frac{    \int _{4m_{c}^2}^{\infty } \left[\sqrt{s} \rho _{\langle \bar{q}q \rangle^3}^1(s)\mp \rho _{\langle \bar{q}q \rangle^3}^0(s) \right] \text{exp}\left(-\frac{s}{M_B^2}\right)ds}
         {\int _{4m_{c}^2}^{\infty} \left[\sqrt{s} \rho ^1(s)\mp \rho^0(s) \right] \text{exp}\left(-\frac{s}{M_B^2}\right)ds} \right\rvert\, ,
            \end{split}
    \end{equation}
\begin{equation}\label{PC}
      \begin{split}
       \text{PC}_{\pm } \equiv \frac{\int _{4m_{c}^2}^{s_0} \left[\sqrt{s} \rho ^1(s)\mp \rho^0(s) \right] \text{exp}\left(-\frac{s}{M_B^2}\right)ds}
         {\int _{4m_{c}^2}^{\infty} \left[\sqrt{s} \rho ^1(s)\mp \rho^0(s) \right] \text{exp}\left(-\frac{s}{M_B^2}\right)ds}\, .
            \end{split}
    \end{equation}    
The CVG is usually required to be small enough to ensure the convergence of OPE series, which provides a lower bound on Borel parameter $M_B^{min}$. The PC is required to be as large as possible to ensure the validity of one-pole parametrization, which can determine the upper limit $M_B^{max}$. 
The continuum threshold value $s_0$ is chosen to minimize the dependence of hadron mass on the  Borel parameter $M_B$. Following these criteria, we plot the variation of hadron mass to $s_0$ for the interpolating current $J^{\Lambda _cD}$ with
$J^P=\frac{1}{2}^-$ (left) and $J^P=\frac{1}{2}^+$ (right) in Fig.~\ref{fig:lanmdc-D-mass-s0}. One can find the optimal working region of $s_0$ around the intersection of the curves with different values of $M_B^2$. The pentaquark mass is almost independent on the Borel parameter $M_B^2$ in this working region of $s_0$. In Fig.~\ref{fig:BorelplatformLambdacD}, we show the their Borel curves which are very stable in the working regions of $s_0$ and $M_B^2$. We obtain the hadron masses for the $\Lambda_cD$ pentaquarks as
\begin{align}
m^{\Lambda_cD}_{1/2^-}=&4.13_{-0.09}^{+0.10}\, \text{GeV}\, ,\\
m^{\Lambda_cD}_{1/2^+}=&4.97_{-0.13}^{+0.14}\, \text{GeV}\, ,
\end{align}
in which the errors come from the uncertainties of the Borel mass $M_B$, the threshold value $s_0$ and various input parameters listed in Eq.~\ref{inputparameter}. One notes that the positive-parity pentaquark is much heavier than the negative-parity state in the same channel since the P-wave excitations. 
\begin{figure}[t!]
  \centering
  \includegraphics[width=8cm]{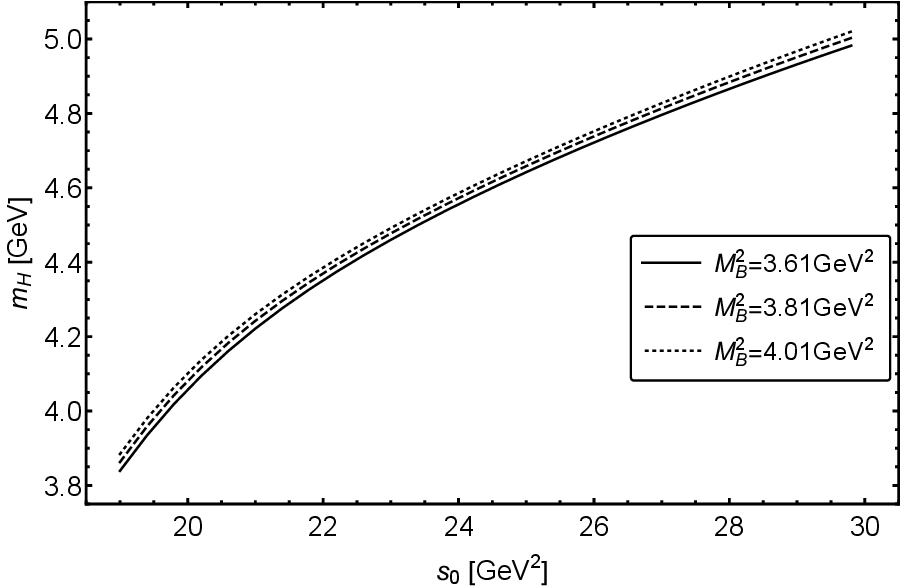}\quad
  \includegraphics[width=8cm]{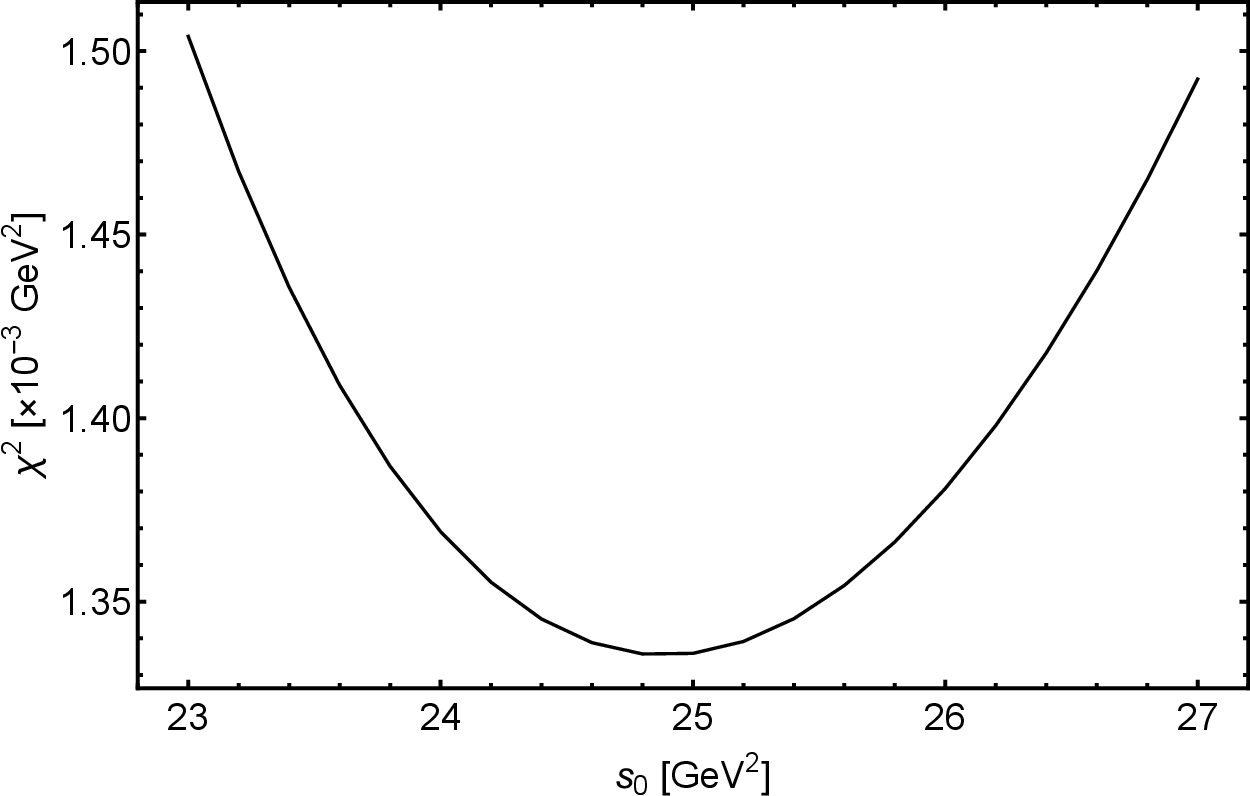}\\
  \caption{Hadron mass (left) and $\chi ^2$ (right) curves for the current $J^{\varSigma _c^*D}$ with $J^P=\frac{3}{2}^+$.}
  \label{fig:sigmc-star-D-fu-mass-s0}
  \end{figure}
\begin{figure}[t!]
  \centering
  \includegraphics[width=10cm]{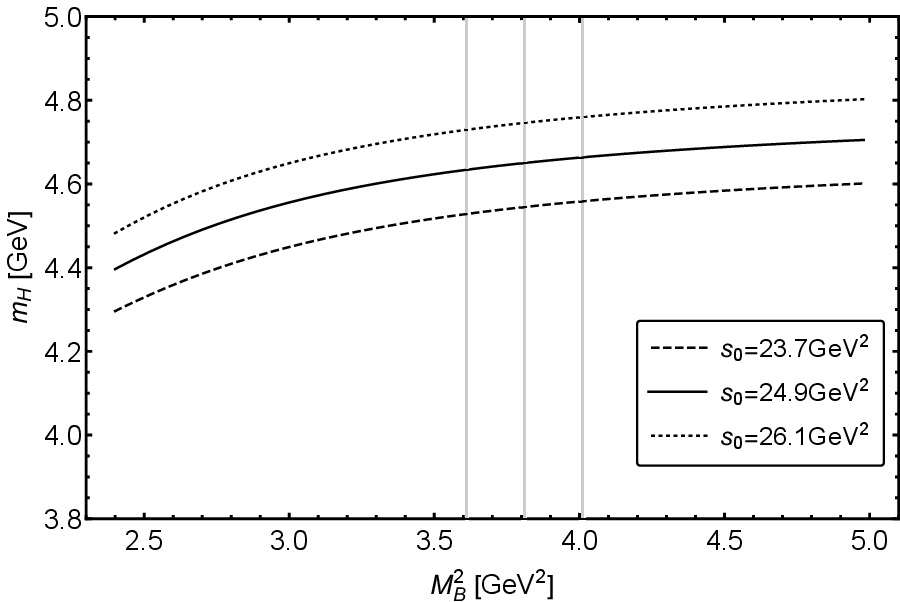}
  \caption{Mass curves for the current $J^{\varSigma _c^*D}$ with $J^P=\frac{3}{2}^+$.}
  \label{fig:BorelplatformSigmacstarD}
  \end{figure}
  
In the left panel of Fig.~\ref{fig:sigmc-star-D-fu-mass-s0}, we show the variation of hadron mass to $s_0$ for the current $J^{\varSigma _c^*D}$ with $J^P=\frac{3}{2}^+$. It is shown that there is no intersection for various values of $M_B^2$ due to the different behavior of the spectral density. To determine the working region of $s_0$ in this situation, we define the following quantity $\chi^2(s_0)$ to study the stability behavior of hadron mass 
      \begin{eqnarray}
    \chi^2(s_0)=\sum_{i,\, j=1; \, i\neq j}^{N}\left[{m_X(s_0,M_{B,i}^2)}-{m_X(s_0,M_{B,j}^2)}\right]^2,
    \end{eqnarray}
in which $M_{B,i} (i=1, 2,..., N)$ represents $N$ definite Borel parameter in its working region. The optimal value of $s_0$ can be obtained by minimizing the quantity of $\chi^2(s_0)$, as shown in the right panel of Fig.~\ref{fig:sigmc-star-D-fu-mass-s0}. The mass curves are plotted in Fig.~\ref{fig:BorelplatformSigmacstarD}, and the corresponding pentaquark mass is predicted as 
\begin{align}
m^{\varSigma _c^*D}_{3/2^+}=&4.64_{-0.17}^{+0.21}\, \text{GeV}\, .
\end{align}

After the numerical analyses, we collect the predicted hadron masses and coupling constants for all negative-parity and positive-parity doubly charmed pentaquarks in Table~\ref{PccfumassTab} and Table~\ref{PcczhmassTab}, respectively. 
We compare the predicted pentaquark masses to the corresponding meson-baryon mass thresholds. In Table~\ref{PccfumassTab}, the charmed baryons $\Lambda _c/\Sigma _c$ and $\Lambda _c^{\ast}/\Sigma _c^{\ast}$ are $J^P=\frac{1}{2}^+$ and $\frac{3}{2}^+$ with positive-parity, respectively. However, we use the P-wave $\Lambda _c(2595)$ and $\Lambda _c^\ast(2625)$ charmed baryons with negative-parity in Table~\ref{PcczhmassTab} to give the two-hadron mass thresholds. As shown in Table~\ref{mesonbaryonTab}, we use the theoretical predicted masses of $\Sigma_c/\Sigma_c^{\ast}$ with $\frac{1}{2}^-/\frac{3}{2}^-$ in the relativized potential quark model~\cite{Capstick:1986ter}. It is shown that the obtained pentaquark masses for the $\Lambda _cD\, (\frac{1}{2}^-)$, $\Sigma _cD\, (\frac{1}{2}^-)$, $\Sigma _cD^*\, (\frac{3}{2}^-)$, $\Lambda _c^*D\, (\frac{3}{2}^-)$, $\Lambda _c^*D^*\, (\frac{5}{2}^-)$ channels with negative-parity and $\Sigma _cD\, (\frac{1}{2}^+)$, $\Sigma _cD^\ast\, (\frac{3}{2}^+)$, $\Sigma _c^\ast D\, (\frac{3}{2}^+)$ channels with positive-parity are lower than the corresponding meson-baryon mass thresholds. The 
doubly charmed $P_{cc}$ pentaquarks in these channels are predicted to form bound states. 

 \begin{table}[h!]
      \caption{ Masses and couplings of the $P_{cc}$ doubly charmed pentaquark  states with negative-parity.}\label{PccfumassTab}
      \renewcommand\arraystretch{1.8} 
      \setlength{\tabcolsep}{0.7 em}{ 
      \begin{tabular}{c c c c c c c c c}
        \hline
         \hline
       current& $J^{P}$ & $s_0 [\mathrm{GeV}^2]$ & $M_B^2 [\mathrm{GeV}^2]$ & Pole & CVG & Mass $[\mathrm{GeV}]$ & Two-hadron hreshold $[\mathrm{GeV}]$& $f_X[\mathrm{GeV^6}]$  \\
        \hline
       $J^{\Lambda _{c}D}$    & $\frac{1}{2}^-$ & $19.5 (\pm 5\%)$ & 2.83$\sim $3.43 & $>16.9\%$ & $<5\%$ & $4.13_{-0.09}^{+0.10}$ & 4.15& $0.77_{-0.16}^{+0.16}\times 10^{-3}$ \vspace{1ex}\\
       $J^{\varSigma _{c}D}$    & $\frac{1}{2}^-$ & $18.3(\pm 5\%)$ & 3.40$\sim $3.70 & $>5.9\%$ & $<5\%$ & $4.08_{-0.13}^{+0.18}$ & 4.32 & $0.28_{-0.08}^{+0.08}\times 10^{-3}$ \vspace{1ex}\\
       $J^{\varSigma _{c}D^*}$    & $\frac{3}{2}^-$ & $20.3(\pm 5\%)$ & 3.17$\sim $3.47 & $>11.9\%$ & $<10\%$ & $4.14_{-0.15}^{+0.18}$ & 4.46 & $0.27_{-0.08}^{+0.08}\times 10^{-3}$ \vspace{1ex}\\
       $J^{\varSigma _{c}^*D}$    & $\frac{3}{2}^-$ & $22.8(\pm 5\%)$ & 3.82$\sim $4.22 & $>13.4\%$ & $<2\%$ & $4.47_{-0.10}^{+0.11}$ & 4.39 & $1.43_{-0.30}^{+0.31}\times 10^{-3}$ \vspace{1ex}\\
       $J^{\Lambda _{c}D^*}$   & $\frac{3}{2}^-$ & $21.0(\pm 5\%)$ & 3.55$\sim $3.95 & $>12.6\%$ & $<5\%$ & $4.31_{-0.10}^{+0.11}$ & 4.29& $0.95_{-0.21}^{+0.21}\times 10^{-3}$ \vspace{1ex}\\
       $J^{\Lambda _{c}^*D}$  & $\frac{3}{2}^-$ & $22.8(\pm 5\%)$ & 2.91$\sim $3.51 & $>25.0\%$ & $<10\%$ & $4.42_{-0.12}^{+0.13}$ & 4.73& $0.79_{-0.15}^{+0.16}\times 10^{-3}$ \vspace{1ex}\\  
       $J^{\Lambda _{c}^*D^*}$  & $\frac{5}{2}^-$ & $22.1(\pm 5\%)$ & 3.09$\sim $3.69 & $>15.5\%$ & $<10\%$ & $4.41_{-0.14}^{+0.17}$ & 4.86 & $0.86_{-0.19}^{+0.21}\times 10^{-3}$ \vspace{1ex}\\   
       $J^{\varSigma _{c}^*D^*}$    & $\frac{5}{2}^-$ & $25.0(\pm 5\%)$ & 4.0$\sim $4.6 & $>12.5\%$ & $<2\%$ & $4.69_{-0.11}^{+0.12}$ & 4.53 & $2.48_{-0.54}^{+0.56}\times 10^{-3}$ \vspace{1ex}\\

       \hline
         \hline
      \end{tabular}
      }
      \end{table}

\begin{table}[h!]
        \caption{Masses and couplings of the $P_{cc}$ doubly charmed pentaquark  states with positive-parity.}\label{PcczhmassTab}
        \renewcommand\arraystretch{1.8} 
        \setlength{\tabcolsep}{0.7 em}{ 
        \begin{tabular}{c c c c c c c c c}
          \hline
           \hline
         current& $J^{P}$ & $s_0 [\mathrm{GeV}^2]$ & $M_B^2 [\mathrm{GeV}^2]$ & Pole & CVG & Mass $[\mathrm{GeV}]$ & Two-hadron threshold $[\mathrm{GeV}]$& $f_X[\mathrm{GeV^6}]$  \\
          \hline
         $J^{\Lambda _{c}D}$    & $\frac{1}{2}^+$ & $27.4(\pm 5\%)$ & 4.11$\sim $4.51 & $>11.7\%$ & $<2\%$ & $4.97_{-0.14}^{+0.13}$ & 4.46 & $1.87_{-0.47}^{+0.49}\times 10^{-3}$ \vspace{1ex}\\
         $J^{\varSigma _{c}D}$    & $\frac{1}{2}^+$ & $23.7(\pm 5\%)$ & 2.98$\sim $3.58 & $>14.8\%$ & $<5\%$ & $4.60_{-0.12}^{+0.11}$ & 4.63 & $0.46_{-0.12}^{+0.12}\times 10^{-3}$ \vspace{1ex}\\
         $J^{\varSigma _{c}D^*}$    & $\frac{3}{2}^+$ & $25.6(\pm 5\%)$ & 4.00$\sim $4.30 & $>11.2\%$ & $<0.5\%$ & $4.77_{-0.09}^{+0.09}$ & 4.77 & $0.49_{-0.12}^{+0.12}\times 10^{-3}$ \vspace{1ex}\\
         $J^{\varSigma _{c}^*D}$    & $\frac{3}{2}^+$ & $24.9(\pm 5\%)$ & 3.61$\sim $4.01 & $>10.9\%$ & $<3\%$ & $4.64_{-0.21}^{+0.17}$ & 4.64 & $0.80_{-0.24}^{+0.25}\times 10^{-3}$ \vspace{1ex}\\
         $J^{\Lambda _{c}D^*}$   & $\frac{3}{2}^+$ & $35.0(\pm 5\%)$ & 4.24$\sim $4.84 & $>27.4\%$ & $<2\%$ & $5.52_{-0.11}^{+0.11}$ & 4.60& $5.32_{-1.08}^{+1.14}\times 10^{-3}$ \vspace{1ex}\\
         $J^{\Lambda _{c}^*D}$  & $\frac{3}{2}^+$ & $22.9(\pm 5\%)$ & 3.06$\sim $3.56 & $>11.4\%$ & $<5\%$ & $4.54_{-0.18}^{+0.15}$ & 4.50 & $0.34 _{-0.10}^{+0.10}\times 10^{-3}$  \vspace{1ex}\\  
         $J^{\Lambda _{c}^*D^*}$  & $\frac{5}{2}^+$ & $33.0(\pm 5\%)$ & 2.99$\sim $3.59 & $>46.4\%$ & $<5\%$ & $5.29 _{-0.14}^{+0.14}$ & 4.64 &  $2.1 _{-0.43}^{+0.44}\times 10^{-3}$  \vspace{1ex}\\   
         $J^{\varSigma _{c}^*D^*}$    & $\frac{5}{2}^+$ & $28.8(\pm 5\%)$ & 3.36$\sim $3.96 & $>17.4\%$ & $<5\%$ & $5.06_{-0.30}^{+0.24}$ & 4.78 & $1.82_{-0.58}^{+0.67}\times 10^{-3}$ \vspace{1ex}\\
         \hline
           \hline
        \end{tabular}
        }
        \end{table}

\section{Summary and discussion}

Motivated by the discoveries of the hidden-charm $P_c/P_{cs}$  pentaquarks, the doubly charmed $\Xi _{cc}^{++}$ and $T_{cc}^+$ states, we have investigated the existence of doubly charmed pentaquark states in the $\Lambda  _{c}^{(*)}D^{(*)}$ and $\Sigma _{c}^{(*)}D^{(*)}$ channels with $J^P=1/2^\pm$, $3/2^\pm$ and $5/2^\pm$ using the QCD sum rule method. We systematically construct the meson-baryon type of interpolating currents for these doubly-charmed pentaquark states and calculate their two-point correlation functions up to dimension-10. We adopt the parity projected sum rules to separate the contributions of negative-parity and positive-parity states, and extract the mass spectra of these $\Lambda _c^{(*)}D^{(*)}$ and $\Sigma _c^{(*)}D^{(*)}$ doubly charmed pentaquarks.

As shown in Table~\ref{PccfumassTab} and Table~\ref{PcczhmassTab}, the negative-parity pentaquarks are much lighter than those with positive-parity in the same channel, which is reasonable since the orbital excitation in the latter systems. It is shown that the doubly charmed pentaquarks in the $\Lambda _cD\, (\frac{1}{2}^-)$, $\Sigma _cD\, (\frac{1}{2}^-)$, $\Sigma _cD^*\, (\frac{3}{2}^-)$, $\Lambda _c^*D\, (\frac{3}{2}^-)$, $\Lambda _c^*D^*\, (\frac{5}{2}^-)$ channels with negative-parity and $\Sigma _cD\, (\frac{1}{2}^+)$, $\Sigma _cD^\ast\, (\frac{3}{2}^+)$, $\Sigma _c^\ast D\, (\frac{3}{2}^+)$ channels with positive-parity lie below the corresponding meson-baryon thresholds, which are predicted to form bound states. However, these doubly charmed $P_{cc}$ states can still decay into $\Xi_{cc}^{(\ast)}$ plus a light meson via the strong interaction, because all of them lie above the $\Xi\pi$ and $\Xi^\ast\pi$ thresholds. Considering the isospin symmetry, we list some possible two-hadron decay modes for the negative-parity and positive-parity $P_{cc}$ states in Table~\ref{tab:decay_mode_Nparity} and Table~\ref{tab:decay_mode_Pparity} respectively, in which the LQCD prediction of the $\Xi_{cc}^\ast$ mass is adopted~\cite{Mondal:2017nhw}. Besides the $\Xi_{cc}^{(\ast)}\pi/\rho/\omega/\eta/\eta^\prime$ modes, the possible charmed baryon-meson final states are also considered. It is especially interesting for triply charged $P_{cc}^{+++}(ccuu\bar d)$ and neutral $P_{cc}^{0}(ccdd\bar u)$ states in the isospin quartet since their exotic charges are different from the ordinary doubly charmed baryons $\Xi_{cc}^{++}/\Xi_{cc}^{+}$. They should be definite pentaquark states if they do exist. One may search for these characteristic signals for doubly charmed pentaquarks in the $P_{cc}^{+++}\to\Xi_{cc}^{(\ast) ++}\pi^+/\rho^+$, $\Sigma_c^{(\ast)++}D^{(\ast)+}$ and $P_{cc}^{0}\to\Xi_{cc}^{(\ast) +}\pi^-/\rho^-$, $\Sigma_c^{(\ast)0}D^{(\ast)0}$ decay processes so long as the kinematics and $I(J^P)$ quantum numbers allow. For the $P_{cc}$ states in the isospin doublet, the $\Lambda_cD^{(\ast)}$ final states are also the important decay modes.

\begin{figure}[t!]
  \centering
  \includegraphics[width=11cm]{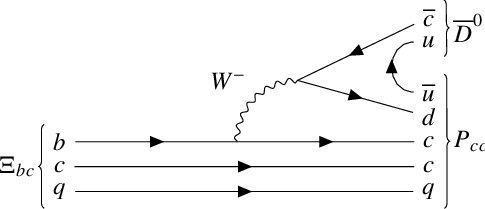}
  \caption{$P_{cc}$ production from the weak decay of doubly heavy $\Xi _{bc}$ baryon.}
  \label{fig:production-Pcc}
  \end{figure}
  
As shown in Fig.~\ref{fig:production-Pcc}, the $P_{cc}$ pentaquark states can be produced via the weak decay of doubly heavy $\Xi _{bc}$ baryon, although it has not been discovered to date~\cite{LHCb:2020iko,LHCb:2021xba,LHCb:2022fbu}. Recently, an inclusive decay channel $\Xi _{bc}\to\Xi_{cc}^{++}+X$~\cite{Qin:2021zqx} and the hadronic production mechanisms at a fixed-target experiment (AFTER@LHC)~\cite{Li:2022mxp} at LHC have been proposed to search for the $\Xi _{bc}$ baryon. Besides, the doubly charmed $P_{cc}$ pentaquarks can also be produced by the weak decays of triply charmed $\Omega _{ccc}$ baryon~\cite{Xing:2021yid}. In the near future, we hope that these doubly charmed pentaquark states can be observed at facilities such as LHCb, BelleII, CMS and RHIC, where copious heavy quarks are produced.

\begin{table}[t!]
	\caption{Some possible two-hadron decay modes of $P_{cc}$ states with negative parity. The notation ``$\varnothing$'' denotes that there is no $P_{cc}$ state with such quantum numbers, and ``$-$'' means no allowed decay mode in the corresponding channel. } 
	\renewcommand\arraystretch{1.3} 
	\setlength{\tabcolsep}{0.7 em}{
		\begin{tabular}{c c c c c }
			\hline
			\hline
			$J^P$ & Current &  Partial wave   & $I=\frac{1}{2}$ & $I=\frac{3}{2}$    \\
			\hline  
			\multirow{4}{*}{$\frac{1}{2}^-$} & \multirow{2}{*}{  $ J^{   \Lambda_c D}$   } &S &  $\Xi_{cc}\pi$  &$ \varnothing $   \\
			& {} & P &   $\Xi_{cc}\sigma$  & $ \varnothing $
			\\ \cline{2-5}
			& \multirow{2}{*}{  $ J^{   \Sigma_c D}$   } &S & $\Xi_{cc}\pi$   & $\Xi_{cc}\pi$     \\ 
			& {} & P &  -  & -  \\
			\hline
			\multirow{8}{*}{$\frac{3}{2}^-$} & \multirow{2}{*}{  $ J^{   \Sigma_c D^* }$   } &S &  $  \Xi_{cc}^* \pi  $ &  $  \Xi_{cc}^* \pi  $   \\
			& {} & P &    $   \Xi_{cc}\sigma   $ & -
			\\  \cline{2-5}
			& \multirow{2}{*}{  $ J^{   \Sigma_c^* D}$   } & S & $ \Lambda_c D^*,  \Sigma_c D^* , \Sigma_c^* D , \Xi_{cc} \rho / \omega ,  \Xi_{cc}^* \pi / \eta $  &  $  \Sigma_c D^* , \Sigma_c^* D , \Xi_{cc} \rho , \Xi_{cc}^* \pi  $   \\ 
			& {} & P &   $  \Lambda_c(2595)D, \Xi_{cc}^{(*)}\sigma  $  & - 
			\\  \cline{2-5}
			& \multirow{2}{*}{  $ J^{   \Lambda_c D^* }$   } &S &  $   \Lambda_c D^*, \Xi_{cc}^* \pi / \eta  $  &$ \varnothing $   \\
			& {} & P &   $  \Xi_{cc}^{(*)}\sigma  $  & $ \varnothing $
			\\  \cline{2-5}
			& \multirow{2}{*}{  $ J^{   \Lambda_c^* D}$   } &S &  $\Lambda_c D^* , \Sigma_c^*D, \Sigma_c^{(*)} D^* , \Xi_{cc}\omega / \rho ,  \Xi_{cc}^*\pi / \omega / \rho / \eta / \eta' $  &$  \varnothing $    \\ 
			& {} & P &   $ \Lambda_c(2595)D^{(*)} ,   \Lambda_c D_0 / D_1 , \Xi_{cc}^{(*)} \sigma / a_0 / f_0(980) $  & $ \varnothing $ 
			\\
			\hline
			\multirow{4}{*}{$\frac{5}{2}^-$} & \multirow{2}{*}{  $ J^{   \Lambda_c^* D^*}$   } &S & -  &$ \varnothing $   \\
			& {} & P &   $\Xi_{cc}^* \sigma$  & $ \varnothing $ 
			\\  \cline{2-5}
			& \multirow{2}{*}{  $ J^{   \Sigma_c^* D^*}$   } &S &  $   \Sigma_c^* D^*,\Xi_{cc}^*\rho / \omega   $  & $ \Sigma_c^* D^*,\Xi_{cc}^*\rho $   \\ 
			& {} & P &   $\Lambda_c(2595)D^* , \Xi_{cc}^* \sigma / f_0(980) / a_0$  & $\Xi_{cc}^* a_0$  \\
			\hline
			\hline
			\label{tab:decay_mode_Nparity}
		\end{tabular}
	}
\end{table}

\begin{table}[t!]
	\caption{Some possible two-hadron decay modes of $P_{cc}$ states with positive parity. The notation ``$\varnothing$'' denotes that there is no $P_{cc}$ state with such quantum numbers, and ``$-$'' means no allowed decay mode in the corresponding channel. } 
	\renewcommand\arraystretch{1.3} 
	\setlength{\tabcolsep}{0.6 em}{
		\begin{tabular}{ccccc}
			\hline
			\hline
			$J^P$ & Current &  Partial wave   & $I=\frac{1}{2}$ & $I=\frac{3}{2}$    \\
			\hline  
			\multirow{4}{*}{$\frac{1}{2}^+$} & \multirow{2}{*}{  $ J^{   \Lambda_c D}$   } &S &  $  \Lambda_c D_0 / D_1  , \Lambda_c (2595) D^{(*)} , \Sigma_c D_0 / D_1     , \Sigma_c^*D_1 , \Xi_{cc} \sigma / f_0(980) / a_0 , \Xi_{cc}^{(*)} a_1   $ &$ \varnothing $   \\
			& {} & P &   $ \Lambda_c D^{(*)} , \Lambda_c(2595)D_0 ,  \Lambda_c^\ast D^{(*)} ,  \Sigma_c^{(*)}D^{(*)} , \Xi_{cc}^{(*)}\pi / \rho /\omega / \eta / \eta'  $  & $ \varnothing $
			\\  \cline{2-5}
			& \multirow{2}{*}{  $ J^{   \Sigma_c D}$   } & S & $  \Lambda_c D_0  , \Lambda_c (2595) D^{(*)} ,   \Xi_{cc} \sigma $   &  -    \\ 
			& {} & P &  $  \Lambda_c D^{(*)} , \Sigma_c^{(*)}D^{(*)} , \Xi_{cc}^{(*)}\pi / \rho /\omega / \eta , \Xi_{cc} \eta' $  &  $  \Sigma_c^{(*)}D^{(*)},   \Xi_{cc}^{(*)}\pi / \rho $  \\
			\hline
			\multirow{8}{*}{$\frac{3}{2}^+$} & \multirow{2}{*}{  $ J^{   \Sigma_c D^* }$   } &S &  $ \Lambda_c D_1 , \Lambda_c(2595)D^* , \Xi_{cc}^* \sigma / f_0(980) / a_0  $ &  -   \\
			& {} & P &    $ \Lambda_c D^{(*)} ,  \Lambda_c^\ast D ,  \Sigma_c^{(*)}D^{(*)} , \Xi_{cc}^{(*)}\pi / \rho / \omega / \eta / \eta'   $ &  $ \Sigma_c^{(*)}D^{(*)} , \Xi_{cc}^{(*)}\pi / \rho  $
			\\   \cline{2-5}
			& \multirow{2}{*}{  $ J^{   \Sigma_c^* D}$   } & S &$ \Lambda_c(2595)D^* , \Xi_{cc}^* \sigma $   &  -  \\ 
			& {} & P &   $ \Lambda_c D^{(*)} , \Sigma_c^{(*)}D^{(*)} , \Xi_{cc}^{(*)}\pi / \rho / \omega / \eta / \eta'   $  & $ \Sigma_c^{(*)}D^{(*)} , \Xi_{cc}^{(*)}\pi / \rho  $ 
			\\   \cline{2-5}
			& \multirow{2}{*}{  $ J^{   \Lambda_c D^* }$   } &S &  $ \Lambda_c D_1 ,  \Lambda_c(2595) D^*, \Sigma_c^* D_0 , \Sigma_c ^{(*)}D_1 ,   \Xi_{cc}^* \sigma / f_0(980) / a_0 ,  \Xi_{cc}^{(*)} a_1 $  &$ \varnothing $   \\
			& {} & P &   $ \Lambda_c D^{(*)} ,  \Lambda_c^\ast D^{(*)} ,\Lambda_c(2595) D_0 / D_1 ,  \Sigma_c^{(*)}D^{(*)} ,  \Xi_{cc}^{(*)}\pi / \rho /\omega / \eta / \eta'  $  & $ \varnothing $
			\\   \cline{2-5}
			& \multirow{2}{*}{  $ J^{   \Lambda_c^* D}$   } &S &  $ \Xi_{cc}^*\sigma  $  &$  \varnothing $    \\ 
			& {} & P &   $  \Lambda_c D^{(*)} , \Sigma_c^{(*)}D^{(*)} ,   \Xi_{cc}^{(*)}\pi / \rho /\omega / \eta   $  & $ \varnothing $ 
			\\
			\hline
			\multirow{4}{*}{$\frac{5}{2}^+$} & \multirow{2}{*}{  $ J^{   \Lambda_c^* D^*}$   } &S & $ \Lambda_c D_2 , \Sigma_c^* D_1 , \Sigma_c^{(*)}D_2 , \Xi_{cc}^* a_1  $ &$ \varnothing $   \\
			& {} & P &   $  \Lambda_c D^* , \Lambda_c^\ast D^{(*)} , \Lambda_c(2595) D_1 / D_2 , \Sigma_c D^*, \Sigma_c^* D^{(*)}  , \Xi_{cc}^{(*)} \rho/\omega , \Xi_{cc}^* \pi  / \eta /\eta' $  & $ \varnothing $ 
			\\    \cline{2-5}
			& \multirow{2}{*}{  $ J^{   \Sigma_c^* D^*}$   } &S & $ \Lambda_c D_2 , \Sigma_c^* D_1 , \Sigma_c^{(*)}D_2 , \Xi_{cc}^* a_1  $ & $\Sigma_c^* D_1 , \Sigma_c^{(*)}D_2 , \Xi_{cc}^* a_1  $   \\ 
			& {} & P &    $  \Lambda_c D^* , \Lambda_c^\ast D^{(*)} , \Lambda_c(2595) D_1 / D_2 , \Sigma_c D^*, \Sigma_c^* D^{(*)}  , \Xi_{cc}^{(*)} \rho/\omega , \Xi_{cc}^* \pi  / \eta /\eta' $ & $ \Sigma_c D^*, \Sigma_c^* D^{(*)}  , \Xi_{cc}^{(*)} \rho , \Xi_{cc}^* \pi  $  \\
			\hline
			\hline
			\label{tab:decay_mode_Pparity}
		\end{tabular}
	}
\end{table}

%
%

\section{ACKNOWLEDGMENTS}
Feng-Bo Duan thanks Peng-Fei Yang and Ding-Kun Lian for valuable discussions. This work is supported by the National Natural Science Foundation of China under Grant No. 12175318 and No. 12305147, the National Key R$\&$D Program of China under Contracts No. 2020YFA0406400, the Natural Science Foundation of Guangdong Province of China under Grant No. 2022A1515011922.


\section*{Appendix: Spectral densities}\label{appendix}

In this appendix, we list the spectral densities $\rho ^{1}(s)$ and $\rho ^{0}(s)$ corresponding to the interpolating currents $J^{\Lambda _{c}^{(*)} D^{(*)}}$ and $J_\mu ^{\varSigma _{c}^{(*)}D^{(*)}}$ separately.
In these expressions, we use the notations
\begin{equation}
\begin{split}
\mathcal{F} (m_c^2)&=m_c^2\left[\frac{1}{x(1-y)}+\frac{1}{y}\right]\, ,\\
x_{min}&=\frac{m_c^2 y}{(1-y)(s y-m_c^2)}\, , ~~~ x_{max}=1\, , \\
y_{min}&=\frac{1-\sqrt {1-4m_c^2/s}}{2}\, , ~~~
y_{max}=\frac{1+\sqrt {1-4m_c^2/s}}{2}\, .
\end{split}
\end{equation}
The spectral densities $\rho ^{1,\Lambda _cD}(s)$ and $\rho ^{0,\Lambda _cD}(s)$ are as follows
{\allowdisplaybreaks
\begin{eqnarray}
  \nonumber\rho _{pert}^{1,\Lambda _cD}(s)&=&\int_{x_{min}}^{x_{max}}  \,dx \int_{y_{min}}^{y_{max}}  \,dy \frac{-1}{(3932160 \pi^8 x^4 y^4)} \{(x-1)^4 (y-1) (m_c^2 (x (-y)+x+y)+s x (y-1) y)^3 (m_c^4 (x (y-1)-y) \\
  \nonumber& & (3 (x-1) y (11 C_{22}-16 x+20)-33 C_{22} x-5 C_{22}+24 x^2+24 (x-1)^2 y^2-60 x-10)-m_c^2 s x (y-1) \\
  \nonumber& & y ((x-1) y (121C_{22}-256 x+260)-121 C_{22} x-5 C_{22}+128 x^2+128 (x-1)^2 y^2-260 x-10)+8 s^2  \\
  \nonumber& & x^2 (y-1)^2 y^2 (11 C_{22}+6 (3 (x-1) y-3 x+5)))\},\\
  \nonumber\rho _{\langle \bar{q}q\rangle}^{1,\Lambda _cD}(s)&=&\langle \bar{q}q\rangle \int_{x_{min}}^{x_{max}}  \,dx \int_{y_{min}}^{y_{max}}  \,dy \frac{-1}{12288 \pi^6 x^3 y^3} \{(x-1)^2 (m_c^2 (x (-y)+x+y)+s x (y-1) y)^2 (m_c^3 (x (y-1)-y) ((11 \\
  \nonumber& & C_{22}+10) x^2 (y-1)^2-x (y-1) (10 (C_{22}+2) y-10 C_{22}-9)-y ((C_{22}-10) y+10 C_{22}+21))-m_c s x (y-\\
  \nonumber& & 1) y (2 (11 C_{22}+10) x^2 (y-1)^2-x (y-1) (20 (C_{22}+2) y-21 C_{22}-19)-y (2 (C_{22}-10) y+9 C_{22}+31)))\},\\
  \nonumber\rho _{\langle GG\rangle}^{1,\Lambda _cD}(s)&=&\langle g_s^2GG\rangle \int_{x_{min}}^{x_{max}}  \,dx \int_{y_{min}}^{y_{max}}  \,dy \frac{1}{23592960 \pi^8 x^4 y^4} \{(x-1)^2 (y-1) (m_c^6 (x (-y)+x+y)^2 (x^5 (y-1)^3 (165 C_{22}\\
  \nonumber& &-212 y+540)-x^4 (y-1)^2 (5 C_{22} (75 y-61)+312 y^2+686 y-670)+x^3 (y-1) (5 C_{22} (38 y (3 y-2)+23)\\
  \nonumber& &+2 y (y (644 y-601)-258)+200)+x^2 (5 C_{22} (3 y (2 y (5 y-22)-1)-5)-2 (y (y (y (636 y-1381)+784)+\\
  \nonumber& &31)+25))+x y (-5 C_{22} (69 y^2+y-2)+2 y (y (134 y-303)-21)+20)-5 y^2 (C_{22} (33 y+5)+12 (5-2 y) y\\
  \nonumber& &+10)+120 x^6 (y-1)^4)-10 m_c^4 s x (y-1) y (x (y-1)-y) (2 x^5 (y-1)^3 (22 C_{22}-23 y+84)-2 x^4 (y-1)^2\\
  \nonumber& & (C_{22} (53 y-42)+2 y (54 y+37)-104)+x^3 (y-1) (C_{22} (y (206 y-105)+36)+2 y (342 y^2-338 y-53)+\\
  \nonumber& &72)-2 x^2 (C_{22} ((132-41 y) y^2+2)+y (2 y (y (174 y-349)+180)+3)+4)+x y (-14 (13 C_{22}+27) y^2+\\
  \nonumber& &(C_{22}-4) y+C_{22}+194 y^3+2)-4 y^2 (11 (C_{22}+2) y+C_{22}-10 y^2+2)+40 x^6 (y-1)^4)+5 m_c^2 s^2\\
  \nonumber& & x^2 (y-1)^2 y^2 (x^5 (y-1)^3 (55 C_{22}+42 y+240)-x^4 (y-1)^2 (C_{22} (155 y-107)+828 y^2-66 y-294)\\
  \nonumber& &+x^3 (y-1) (C_{22} (2 y (230 y-73)+49)+2 y (3 y (362 y-373)+19)+108)+x^2 (C_{22} (y (20 y (13 y-33)\\
  \nonumber& &+9)-3)+2 y (y (3 (713-378 y) y-1010)+7)-6)+x y^2 (-565 C_{22} y+9 C_{22}+762 y^2-1344 y+\\
  \nonumber& &14)+y^2 (-C_{22} (55 y+3)+60 (y-2) y-6)+60 x^6 (y-1)^4)+2 s^3 x^4 (y-1)^3 y^4 (15 C_{22} (x (3 (x-8) y\\
  \nonumber& &-3 x+2)-45 y+1)-259 x^3 (y-1)^2+x^2 (y-1) (917 y-622)+x ((2224-1757 y) y-397)+y \\
  \nonumber& &(1099 y-1803)+34))\},\\
  \nonumber\rho _{\langle \bar{q} Gq\rangle}^{1,\Lambda _cD}(s)&=&\langle g_s \bar{q}\sigma Gq\rangle \int_{x_{min}}^{x_{max}}  \,dx \int_{y_{min}}^{y_{max}}  \,dy \Bigg\{ \frac{1}{16384 \pi^6 x^2 y^2} \{(x-1) (x (y-1) (m_c^2-s y)-m_c^2 y) (m_c^3 (x (y-1)-y) (3\\
  \nonumber& & (11 C_{22}+10) x^2 (y-1)^2-x (y-1) (30 (C_{22}+2) y-31 C_{22}-28)-y (3 (C_{22}-10) y+19 C_{22}+52))\\
  \nonumber& &-m_c s x (y-1) y (5 (11 C_{22}+10) x^2 (y-1)^2-x (y-1) (50 (C_{22}+2) y-53 C_{22}-48)-y (5 (C_{22}-10) y\\
  \nonumber& &+17 C_{22}+72)))\}+\frac{1}{147456 \pi^6 x^3 y^3} \{(x-1) (m_c^2 (x (-y)+x+y)+s x (y-1) y) (m_c s x (y-1) y (-x^3 (y-1)^2\\
  \nonumber& & (2 C_{22} (125 y-9)+325 y-78)+x^2 (y-1) (C_{22} (y (230 y-281)+18)-y (40 y+383)+48)+x y (C_{22}\\
  \nonumber& & (5 y (4 y-41)-7)+y (665 y-823)-37)+66 y^2 (3 C_{22}-5 y+8)+30 x^4 (y-1)^3)-3 m_c^3 (x (y-1)-\\
  \nonumber& &y) (-x^3 (y-1)^2 (C_{22} (50 y-6)+65 y-18)+x^2 (y-1) (C_{22} (y (46 y-61)+6)-y (8 y+81)+12)+x y\\
  \nonumber& & (C_{22} (y (4 y-65)-3)+7 y (19 y-27)-9)+66 y^2 (C_{22}-y+2)+6 x^4 (y-1)^3))\}\Bigg\} ,\\
  \nonumber\rho _{\langle \bar{q} q\rangle^2}^{1,\Lambda _cD}(s)&=&\langle \bar{q} q\rangle^2 \int_{x_{min}}^{x_{max}}  \,dx \int_{y_{min}}^{y_{max}}  \,dy  \frac{1}{3072 \pi^4 x y} \{(x-1) (y-1) (m_c^2 (x (-y)+x+y)+s x (y-1) y) ((13 C_{22}+2) x (y\\
  \nonumber& &-1) (5 s y-3 m_c^2)+m_c^2 (C_{22} (39 y+46)+6 y+44)) \},\\
  \nonumber\rho _{\langle \bar{q} q\rangle\langle \bar{q}G q\rangle}^{1,\Lambda _cD}(s)&=&\langle \bar{q} q\rangle \langle g_s^2 \bar{q}\sigma G q\rangle \int_{x_{min}}^{x_{max}}  \,dx \int_{y_{min}}^{y_{max}}  \,dy\Bigg\{  \frac{1}{3072 \pi^4} \{(y-1) ((13 C_{22}+2) x (y-1) (4 s y-3 m_c^2)+m_c^2 (C_{22} (39 y\\
  \nonumber& &+23)+6 y+22))\}+\frac{-1}{6144 \pi^4 x y} \{(y-1) (m_c^2 (-3 (12 C_{22}+1) x^3 (y-1)^2+x^2 (y-1) (C_{22} (69 y-39)+3 y\\
  \nonumber& &-7)-x (C_{22} (y (33 y-5)+3)+3 y (y+8)+4)+y (45 C_{22}+3 y+44))+4 s x (y-1) y ((12 C_{22}+1) x^2\\
  \nonumber& & (y-1)+C_{22} x (12-11 y)+x+y))\}\Bigg\},\\
  \nonumber\rho _{\langle \bar{q}G q\rangle^2}^{1,\Lambda _cD}(s)&=&\langle g_s^2 \bar{q}\sigma G q\rangle ^2 \int_{x_{min}}^{x_{max}}  \,dx \int_{y_{min}}^{y_{max}}  \,dy  \Bigg\{\frac{1}{49152 \pi^4} \{(y-1) (C_{22} (17 x (y-1)-14 y+13)+17 x (y-1)-17 y+12)\}\\
  \nonumber& &+\frac{-1}{8192 \pi^4} \{(y-1) ((12 C_{22}+1) x (y-1)-y)\}\Bigg\}+\langle g_s^2 \bar{q}\sigma G q\rangle ^2 \int_{0}^{1}  \,dx \int_{0}^{1}  \,dy\Bigg\{ \frac{1}{147456 \pi^4 x y} \{m_c^2 (C_{22} (-17 x^2\\
  \nonumber& & (y-1)^2+x (31 y+2) (y-1)+(145-14 y) y+6)-17 (x (-y)+x+y)^2+147 y)\} +\frac{1}{24576 \pi^4 x y} \{m_c^2 ((12\\
\nonumber& & C_{22}+1) x^2 (y-1)^2-x (y-1) (12 C_{22} y+3 C_{22}+2 y+4)+y (-45 C_{22}+y-44))\}\Bigg\}\delta (s-\mathcal{F} (m_c^2)),\\
 \rho _{\langle \bar{q} q\rangle^3}^{1,\Lambda _cD}(s)&=&\langle \bar{q} q\rangle ^3 \int_{x_{min}}^{x_{max}}  \,dx   \frac{1}{1152 \pi^2} \{C_{22} (m_c-12 m_c x) \},
\end{eqnarray}
}

{\allowdisplaybreaks
\begin{eqnarray}
  \nonumber\rho _{pert}^{0,\Lambda _cD}(s)&=&\int_{x_{min}}^{x_{max}}  \,dx \int_{y_{min}}^{y_{max}}  \,dy \frac{1}{7864320 \pi^8 x^5 y^5} \{C_{22} m_c (x-1)^4 (11 x (y-1)+y) (m_c^2 (x (-y)+x+y)+s x (y-1) y)^4 \\
  \nonumber& & (2 m_c^2 (x (-y)+x+y)+7 s x (y-1) y)\},\\
  \nonumber\rho _{\langle \bar{q}q\rangle}^{0,\Lambda _cD}(s)&=&\langle \bar{q}q\rangle \int_{x_{min}}^{x_{max}}  \,dx \int_{y_{min}}^{y_{max}}  \,dy \frac{1}{49152 \pi^6 x^3 y^3}\{(x-1)^2 (m_c^2 (x (-y)+x+y)+s x (y-1) y)^2 (m_c^4 y (y (-20 (C_{22}+1) y\\
  \nonumber& &+23 C_{22}+22)+26 C_{22}+4)+2 m_c x (y-1) (m_c^3 (30 (C_{22}+1) y^2-(23 C_{22}+22) y-13 C_{22}-2)+m_c s y  \\
  \nonumber& &(-60 (C_{22}+1) y^2+66 C_{22} y+13 C_{22}+64 y+2))+20 (C_{22}+1) x^3 (y-1)^3 (m_c^4-6 m_c^2 s y+7 s^2 y^2)+x^2 \\
  \nonumber& & (y-1)^2 (m_c^4 (-60 (C_{22}+1) y+23 C_{22}+22)+4 m_c^2 s y (60 (C_{22}+1) y-33 C_{22}-32)+s^2 y^2 (-140 (C_{22}\\
  \nonumber& & +1) y +149 C_{22}+146)))\},\\
  \nonumber\rho _{\langle GG\rangle}^{0,\Lambda _cD}(s)&=&\langle g_s^2GG\rangle \int_{x_{min}}^{x_{max}}  \,dx \int_{y_{min}}^{y_{max}}  \,dy \frac{1}{9437184 \pi^8 x^5 y^5} \{C_{22} (x-1)^2 (2 m_c^7 (x (-y)+x+y)^3 (22 x^5 (y-1)^3+x^4 (y-1)^2 (9 y\\
  \nonumber& &+44)-11 x^3 ((13 y-15) y^2+2)+x^2 (y-1) y (23 y-13)+x y^2 (15 y-11)+2 y^3)+3 m_c^5 s x (y-1) y (x (-y)+x\\
  \nonumber& &+y)^2 (55 x^5 (y-1)^3+x^4 (y-1)^2 (37 y+110)-x^3 (y-1) (428 y^2-64 y-55)+x^2 y (4 y (20 y-27)+27)+7 x y^2\\
  \nonumber& & (5 y-3)+5 y^3)-4 m_c^3 s^2 x^2 (y-1)^2 y^2 (x (y-1)-y) (44 x^5 (y-1)^3+x^4 (y-1)^2 (47 y+88)-x^3 (y-1) (y (427 y\\
  \nonumber& &-62)-44)+x^2 y (y (91 y-108)+15)+x y^2 (25 y-9)+4 y^3)+5 m_c s^3 x^3 (y-1)^3 y^3 (11 x^5 (y-1)^3+x^4 (y-1)^2 \\
  \nonumber& &(19 y+22)-x^3 (y-1) (2 y (71 y-10)-11)+x^2 y (34 y^2-36 y+1)+x y^2 (5 y+1)+y^3))\},\\
  \nonumber\rho _{\langle \bar{q} Gq\rangle}^{0,\Lambda _cD}(s)&=&\langle g_s \bar{q}\sigma Gq\rangle \int_{x_{min}}^{x_{max}}  \,dx \int_{y_{min}}^{y_{max}}  \,dy \Bigg\{\frac{-1}{49152 \pi^6 x^2 y^2} \{(x-1) (x (y-1) (m_c^2-s y)-m_c^2 y) (m_c^4 y (2 y (-30 (C_{22}+1) y\\
  \nonumber& &+33 C_{22}+32)+39 C_{22}+6)+m_c x (y-1) (m_c^3 (180 (C_{22}+1) y^2-4 (33 C_{22}+32) y-39 C_{22}-6)\\
  \nonumber& &+m_c s y (-300 (C_{22}+1) y^2+(321 C_{22}+314) y+39 C_{22}+6))+60 (C_{22}+1) x^3 (y-1)^3 (m_c^4-5 m_c^2 s y\\
  \nonumber& &+5 s^2 y^2)+x^2 (y-1)^2 (2 m_c^4 (-90 (C_{22}+1) y+33 C_{22}+32)+m_c^2 s y (600 (C_{22}+1) y-321 C_{22}-314)\\
  \nonumber& &-5 s^2 y^2 (60 (C_{22}+1) y-63 C_{22}-62)))\}+\frac{1}{294912 \pi^6 x^3 y^3} \{(x-1) (x (y-1) (m_c^2-s y)-m_c^2 y) (m_c^4 (64 (C_{22}\\
  \nonumber& &+1) x^5 (y-1)^4-2 x^4 (y-1)^3 (4 C_{22} (7 y-17)+40 y-65)-x^3 (y-1)^2 (C_{22} (4 y (56 y+51)-81)+8 y (19 y\\
  \nonumber& &+25)-57)+x^2 (y-1) (8 (46 C_{22}+37) y^3-4 (2 C_{22}+1) y^2+27 (C_{22}-5) y+9 (C_{22}-1))+x y (C_{22} (y (4 \\
  \nonumber& &(21-40 y) y-315)+198)+y (8 (11-17 y) y+63)+18)+y^2 (8 (C_{22}+1) y^2+207 C_{22}-6 y+9))\\
  \nonumber& &+m_c^2 s x (y-1) y (-320 (C_{22}+1) x^4 (y-1)^3-x^3 (y-1)^2 (40 (C_{22}-2) y+668 C_{22}+647)+x^2 (y-1) (120\\
  \nonumber& & (9 C_{22}+7) y^2+(310 C_{22}+353) y-357 C_{22}-318)+x (-40 (19 C_{22}+16) y^3+(410 C_{22}+367) y^2\\
  \nonumber& &+156 (C_{22}+2) y-9 C_{22}+9)+y (40 (C_{22}+1) y^2-3 (4 C_{22}+11) y+207 C_{22}+9))+5 s^2 x^2 (y-1)^2 y^2\\
  \nonumber& & (64 (C_{22}+1) x^3 (y-1)^2+3 x^2 (y-1) (8 (3 C_{22}+2) y+44 C_{22}+43)+x (-24 (6 C_{22}+5) y^2+74 C_{22} y\\
  \nonumber& &+68 C_{22}+58 y+65)+y (8 (C_{22}+1) y-4 C_{22}-7)))\}\Bigg\},\\
  \nonumber\rho _{\langle \bar{q} q\rangle^2}^{0,\Lambda _cD}(s)&=&\langle \bar{q} q\rangle^2 \int_{x_{min}}^{x_{max}}  \,dx \int_{y_{min}}^{y_{max}}  \,dy  \frac{-1}{768 \pi^4 x^2 y^2} \{m_c (x-1) ((10 C_{22}+21) x (y-1)-(10 C_{22}+9) y) (m_c^2 (x (-y)+x\\
  \nonumber& &+y) +s x (y-1) y) (m_c^2 (x (-y)+x+y)+2 s x (y-1) y) \},\\
  \nonumber\rho _{\langle \bar{q} q\rangle\langle \bar{q}G q\rangle}^{0,\Lambda _cD}(s)&=&\langle \bar{q} q\rangle \langle g_s^2 \bar{q}\sigma G q\rangle \int_{x_{min}}^{x_{max}}  \,dx \int_{y_{min}}^{y_{max}}  \,dy\Bigg\{  \frac{1}{1536 \pi^4 x y} \{((10 C_{22}+21) m_c x (y-1)-(10 C_{22}+9) m_c y) (x (y-1) (2 m_c^2\\
  \nonumber& &-3 s y)-2 m_c^2 y)\}+\frac{1}{6144 \pi^4 x^2 y^2} \{m_c ((5 C_{22}-17) x^3 (y-1)^2+x^2 (y-1) (C_{22} (y+5)+22 y-17)+x y (C_{22} (6\\
  \nonumber& &-37 y)-34 y+4)+(43 C_{22}+41) y^2) (2 m_c^2 (x (y-1)-y)-3 s x (y-1) y)\}\Bigg\},\\
  \nonumber\rho _{\langle \bar{q}G q\rangle^2}^{0,\Lambda _cD}(s)&=&\langle g_s^2 \bar{q}\sigma G q\rangle ^2 \int_{x_{min}}^{x_{max}}  \,dx \int_{y_{min}}^{y_{max}}  \,dy  \Bigg\{\frac{-1}{24576 \pi^4 x y} \{(C_{22}+1) m_c (3 x^2 (y-1)^2+x (37 y+6) (y-1)+y (3 y+43))\}\\
  \nonumber& &+\frac{1}{12288 \pi^4 x y} \{m_c (-(5 C_{22}-17) x^2 (y-1)^2-2 (3 C_{22}+2) x y (y-1)+(43 C_{22}+41) y^2)\}\Bigg\}+\langle g_s^2 \bar{q}\sigma G q\rangle ^2  \\
  \nonumber& &\int_{0}^{1}  \,dx \int_{0}^{1}  \,dy\Bigg\{ \frac{1}{49152 \pi^4 x^2 (y-1) y^2} \{(C_{22}+1) m_c^3 (x (y-1)-y) (3 x^2 (y-1)^2+x (37 y+6) (y-1)+y (3 y\\
  \nonumber& &+43))\} +\frac{1}{24576 \pi^4 x^2 (y-1) y^2} \{m_c^3 (x (y-1)-y) ((5 C_{22}-17) x^2 (y-1)^2+2 (3 C_{22}+2) x y (y-1)-\\
  \nonumber& &(43 C_{22}+41) y^2)\}\Bigg\}\delta (s-\mathcal{F} (m_c^2)),\\
  \rho _{\langle \bar{q} q\rangle^3}^{0,\Lambda _cD}(s)&=&\langle \bar{q} q\rangle ^3 \int_{x_{min}}^{x_{max}}  \,dx   \frac{1}{576 \pi^2} \{(C_{22}+2) (13 m_c^2+3 s (x-1) x) \},
\end{eqnarray}
}
where the parameter $C_{22}=1$ and $C_{22}=-1$ for the currents $J^{\Lambda_c D}$ and $J^{\Sigma_c D}$, respectively. 
The spectral densities $\rho ^{1,\Lambda _cD^*}(s)$ and $\rho ^{0,\Lambda  _cD^*}(s)$ are as follows
{\allowdisplaybreaks
\begin{eqnarray}
  \nonumber\rho _{pert}^{1,\Lambda _cD^*}(s)&=&\int_{x_{min}}^{x_{max}}  \,dx \int_{y_{min}}^{y_{max}}  \,dy \frac{-1}{19660800 \pi^8 x^4 y^4} \{C_{11} (x-1)^4 (y-1) (m_c^2 (x (-y)+x+y)+s x (y-1) y)^3 (C_{21} (m_c^4 (x\\
  \nonumber& & (y-1)-y) (76 x^2 (y-1)^2-x (152 y-211) (y-1)+y (76 y-211)-45)-m_c^2 s x (y-1) y (497 x^2 (y-1)^2\\
  \nonumber& &-2 x (497 y-531) (y-1)+y (497 y-1062)-45)+s^2 x^2 (y-1)^2 y^2 (621 x (y-1)-621 y+1051))+25 C_{22} \\
  \nonumber& &(m_c^2 ((x-1) y-x)-s x (y-1) y) (5 m_c^2 ((x-1) y-x)-17 s x (y-1) y))\},\\
  \nonumber\rho _{\langle \bar{q}q\rangle}^{1,\Lambda _cD^*}(s)&=&\langle \bar{q}q\rangle \int_{x_{min}}^{x_{max}}  \,dx \int_{y_{min}}^{y_{max}}  \,dy \frac{-1}{36864 \pi^6 x^3 y^3} \{C_{11} m_c (x-1)^2 (m_c^2 (x (-y)+x+y)+s x (y-1) y)^2 (x^3 (y-1)^3 (m_c^2 (\\
  \nonumber& &19 C_{21}+24 C_{22})-s y (49 C_{21}+60 C_{22}))+x^2 (y-1)^2 (C_{21} m_c^2 (16-69 y)+2 C_{21} s y (55 y-23)+C_{22} m_c^2 \\
  \nonumber& &(24-47 y)+C_{22} s y (59 y-60))+x y (y-1) (C_{21} m_c^2 (81 y-80)+C_{21} s (94-61 y) y+C_{22} m_c^2 (22 y-62)\\
  \nonumber& &+C_{22} s y (y+38))+m_c^2 y^2 (C_{21} (64-31 y)+C_{22} (y+38)))\},\\
  \nonumber\rho _{\langle GG\rangle}^{1,\Lambda _cD^*}(s)&=&\langle g_s^2GG\rangle \int_{x_{min}}^{x_{max}}  \,dx \int_{y_{min}}^{y_{max}}  \,dy \frac{1}{283115520 \pi^8 x^4 y^4} \{C_{11} (x-1)^2 (y-1) (5 C_{22} (m_c^2 ((x-1) y-x)-s x (y-1) y) ((\\
  \nonumber& &3 (y-1)^4 (3 y+100) x^6-(y-1)^3 (y (36 y+883)-600) x^5+(y-1)^2 (y (54 y^2+3340 y-1195)+300) x^4-\\
  \nonumber& &3 (y-1) y (2 y (2 y (3 y+524)-221)+93) x^3+y (y (y (y (9 y+4888)-2067)-66)+24) x^2+y^2 ((279-1657 y) y\\
  \nonumber& &+24) x+300 y^4) m_c^4-s x (y-1) y (3 (y-1)^3 (9 y+196) x^5-(y-1)^2 (y (81 y+1121)-1176) x^4+(y-1) (y (9 y\\
  \nonumber& & (9 y+959)-1157)+588) x^3+y (y (2059-3 y (9 y+3593))+15) x^2+y (y (3269 y+15)-24) x-588 y^3) m_c^2\\
  \nonumber& &+2 s^2 x^3 (y-1)^2 y^3 (-668 y+x (9 x^2 (y-1)^2-x (18 y-19) (y-1)+y (9 y+2812)+7)+3))+2 C_{21} ((-y x+x\\
  \nonumber& &+y)^2 (456 (y-1)^4 x^6-9 (y-1)^3 (21 y-242) x^5+6 (y-1)^2 (y (27 y-212)+453) x^4-(y-1) (y (6 y (628 y-\\
  \nonumber& &901)+425)-726) x^3+2 (y (y (3 y (997 y-2271)+32)+275)-135) x^2+y (y (-3099 y^2+6393 y+658)\\
  \nonumber& &+108) x+6 y^2 (y (76 y-211)-45)) m_c^6-3 s x (x (y-1)-y) (y-1) y (580 (y-1)^4 x^6+(y-1)^3 (405 y+2516) x^5\\
  \nonumber& &-4 (y-1)^2 (y (165 y-64)-787) x^4-(y-1) (y (5490 y^2-5992 y-361)-1068) x^3+2 (y (y (4510 y^2-9396 y\\
  \nonumber& &+251)+237)-72) x^2+y (y ((8109-4435 y) y+510)+36) x+4 y^2 (y (145 y-339)-36)) m_c^4+3 s^2 x^2 (y-1)^2\\
  \nonumber& & y^2 (468 (y-1)^4 x^6+5 (y-1)^3 (395 y+382) x^5-2 (y-1)^2 (31 y (41 y-68)-1181) x^4-(y-1) (y (7052 y^2-5202 y\\
  \nonumber& &-2647)-866) x^3+2 (y (y (y (5979 y-11771)+426)+203)-27) x^2+y^2 ((8849-5275 y) y+406) x+2 y^2 (y\\
  \nonumber& & (234 y-487)-27)) m_c^2-s^3 x^4 (y-1)^3 y^4 (4881 (y-1)^2 x^3-3 (y-1) (321 y-3776) x^2+(3 (4668-3059 y) y\\
  \nonumber& &+6793) x+5259 y^2-8253 y-346)))\},\\
  \nonumber\rho _{\langle \bar{q} Gq\rangle}^{1,\Lambda _cD^*}(s)&=&\langle g_s \bar{q}\sigma Gq\rangle \int_{x_{min}}^{x_{max}}  \,dx \int_{y_{min}}^{y_{max}}  \,dy \Bigg\{ \frac{-1}{589824 \pi^6 x^3 y^2} \{C_{11} m_c (x-1) (m_c^2 (x (y-1)-y)-s x (y-1) y) (2 C_{21} x\\
  \nonumber& & (m_c^2 (x (y-1)-y) (175 x^2 (y-1)^2-x (134 y-149) (y-1)+(67-41 y) y-32)-s x (y-1) y (355 x^2 (y\\
  \nonumber& &-1)^2-x (278 y-317) (y-1)+(115-77 y) y-44))+C_{22} (m_c^2 (x (y-1)-y) (440 x^3 (y-1)^2-x^2 (y\\
  \nonumber& &-1) (448 y-439)+x (y (20 y+29)-1)-12 y (y+2))+s x y (-872 x^3 (y-1)^3+x^2 (880 y-871) (y-1)^2\\
  \nonumber& &-x (y (20 y+29)-1) (y-1)+12 y (y^2+y-2))))\}+\frac{1}{16384 \pi^6 x^2 y^2} \{C_{11} m_c (x-1) (x (y-1) (m_c^2-s y)\\
  \nonumber& &-m_c^2 y) (x^3 (y-1)^3 (m_c^2 (19 C_{21}+24 C_{22})-3 s y (13 C_{21}+16 C_{22}))+x^2 (y-1)^2 (C_{21} m_c^2 (17-69 y)+\\
  \nonumber& &C_{21} s y (90 y-37)+C_{22} m_c^2 (24-47 y)+C_{22} s y (47 y-48))+x y (y-1) (C_{21} m_c^2 (81 y-70)+C_{21} s (73\\
  \nonumber& &-51 y) y+C_{22} m_c^2 (22 y-49)+C_{22} s y (y+25))+m_c^2 y^2 (C_{21} (53-31 y)+C_{22} (y+25)))\}\Bigg\} ,\\
  \nonumber\rho _{\langle \bar{q} q\rangle^2}^{1,\Lambda _cD^*}(s)&=&\langle \bar{q} q\rangle^2 \int_{x_{min}}^{x_{max}}  \,dx \int_{y_{min}}^{y_{max}}  \,dy  \frac{1}{1536 \pi^4 x y} \{C_{11} (x-1) (y-1) (m_c^2 (x (-y)+x+y)+s x (y-1) y) (x (y-1) (s y \\
  \nonumber& &(5 C_{21}+24 C_{22})-3 m_c^2 (C_{21}+4 C_{22}))+m_c^2 (C_{21} (3 y+22)+12 C_{22} (y+2))) \},\\
  \nonumber\rho _{\langle \bar{q} q\rangle\langle \bar{q}G q\rangle}^{1,\Lambda _cD^*}(s)&=&\langle \bar{q} q\rangle \langle g_s^2 \bar{q}\sigma G q\rangle \int_{x_{min}}^{x_{max}}  \,dx \int_{y_{min}}^{y_{max}}  \,dy\Bigg\{  \frac{1}{1536 \pi^4} \{C_{11} (y-1) (x (y-1) (2 s y (2 C_{21}+9 C_{22})-3 m_c^2 (C_{21}\\
  \nonumber& &+4 C_{22}))+m_c^2 (C_{21} (3 y+11)+12 C_{22} (y+1)))\}+\frac{1}{36864 \pi^4 x y} \{C_{11} (y-1) (2 C_{21} m_c^2 (-11 x^3 (y-1)^2\\
  \nonumber& &+x^2 (y-1) (53 y-11)-x y (53 y+8)+11 y^2)+4 C_{21} s x (y-1) y (x (8 x (y-1)-29 y+8)+8 y)+\\
  \nonumber& &C_{22} m_c^2 (138 x^3 (y-1)^2-6 x^2 (y-1) (43 y-23)+x y (108 y-199)+12 y^2)-C_{22} s x (y-1) y (x (211 \\
  \nonumber& &x (y-1)-191 y+211)-14 y))\}\Bigg\},\\
  \nonumber\rho _{\langle \bar{q}G q\rangle^2}^{1,\Lambda _cD^*}(s)&=&\langle g_s^2 \bar{q}\sigma G q\rangle ^2 \int_{x_{min}}^{x_{max}}  \,dx \int_{y_{min}}^{y_{max}}  \,dy  \Bigg\{\frac{1}{884736 \pi^4} \{C_{11} (y-1) (C_{21} (-86 x (y-1)+86 y-56)+C_{22} (13 x (y\\
  \nonumber& &-1)+45 y+2))\}+\frac{1}{73728 \pi^4} \{C_{11} (y-1) (x (y-1) (11 C_{21}-69 C_{22})-y (11 C_{21}+6 C_{22}))\}\Bigg\}\\
  \nonumber& &+\langle g_s^2 \bar{q}\sigma G q\rangle ^2 \int_{0}^{1}  \,dx \int_{0}^{1}  \,dy\Bigg\{ \frac{1}{1769472 \pi^4 x y} \{C_{11} m_c^2 (12 C_{21} (7 (x-1)^2 y^2+2 (8-7 x) x y+x (7 x-2)\\
  \nonumber& &+y-2)+C_{22} (-4 x^2 (y-1)^2-x (39 y+2) (y-1)+y (43 y-12))) \}+\frac{-1}{147456 \pi^4 x y} \{C_{11} m_c^2 (x (y-1)\\
  \nonumber& &-y) (x (y-1) (10 C_{21}-73 C_{22})-2 y (5 C_{21}+C_{22}))\}\Bigg\}\delta (s-\mathcal{F} (m_c^2)),\\
 \rho _{\langle \bar{q} q\rangle^3}^{1,\Lambda _cD^*}(s)&=&\langle \bar{q} q\rangle ^3 \int_{x_{min}}^{x_{max}}  \,dx   \frac{-1}{576 \pi^2} 5 C_{11} C_{22} m_c x,
\end{eqnarray}
}

{\allowdisplaybreaks
\begin{eqnarray}
  \nonumber\rho _{pert}^{0,\Lambda _cD^*}(s)&=&\int_{x_{min}}^{x_{max}}  \,dx \int_{y_{min}}^{y_{max}}  \,dy \frac{1}{19660800 \pi^8 x^5 y^5} \{C_{11} C_{22} (x-1)^4 (y-1) (m_c^2 (x (-y)+x+y)+s x (y-1) y)^4 (m_c s x (y-1) \\
  \nonumber& &y (7 (x-1)^2 y+(187-7 x) x)-2 m_c^3 (x (y-1)-y) (x ((x-2) y-x+16)+y))\},\\
  \nonumber\rho _{\langle \bar{q}q\rangle}^{0,\Lambda _cD^*}(s)&=&\langle \bar{q}q\rangle \int_{x_{min}}^{x_{max}}  \,dx \int_{y_{min}}^{y_{max}}  \,dy \frac{1}{147456 \pi^6 x^3 y^3} \{C_{11} (x-1)^2 (m_c^2 (x (-y)+x+y)+s x (y-1) y)^2 (m_c^3 y (2 C_{21} m_c (2-7 \\
  \nonumber& &(y-2) y)+C_{22} m_c ((59-56 y) y+72))+2 m_c^2 x (y-1) (C_{21} m_c^2 (7 y (3 y-4)-2)+2 C_{21} s y ((63-55 y) y+1)\\
  \nonumber& &+C_{22} m_c^2 (y (84 y-59)-36)+C_{22} s y ((187-184 y) y+36))+2 x^3 (y-1)^3 (7 m_c^4 (C_{21}+4 C_{22})-2 m_c^2 s y (55 \\
  \nonumber& &C_{21}+92 C_{22})+s^2 y^2 (163 C_{21}+228 C_{22}))-x^2 (y-1)^2 (2 C_{21} (7 m_c^4 (3 y-2)+2 m_c^2 s (63-110 y) y+s^2 y^2 \\
  \nonumber& &(163 y-172))+C_{22} (m_c^4 (168 y-59)+2 m_c^2 s (187-368 y) y+3 s^2 y^2 (152 y-153))))\},\\
  \nonumber\rho _{\langle GG\rangle}^{0,\Lambda _cD^*}(s)&=&\langle g_s^2GG\rangle \int_{x_{min}}^{x_{max}}  \,dx \int_{y_{min}}^{y_{max}}  \,dy \frac{1}{141557760 \pi^8 x^5 y^5} \{C_{11} C_{22} (x-1)^2 (m_c^7 (x (-y)+x+y)^3 (24 x^6 (y-1)^4-4 x^5 (y\\
  \nonumber& &-1)^3 (5 y-108)+x^4 (y-1)^2 (y (138 y+137)+792)-2 x^3 (y-1) (y (74 y^2+442 y-177)-192)+x^2 y (2 y(y \\
  \nonumber& & (3 y-499)+280)+197)+x y^2 ((271-24 y) y-177)+24 (y-1) y^3)+3 m_c^5 s x (y-1) y (x (-y)+x+y)^2 (30 \\
  \nonumber& &x^6 (y-1)^4-6 x^5 (y-1)^3 (2 y-115)+3 x^4 (y-1)^2 (9 y (7 y+17)+430)-x^3 (y-1) (y (y (210 y+2239)-734)\\
  \nonumber& &-630)+x^2 y (y (775-y (9 y+1264))+263)-2 x y^2 (y (9 y-163)+119)+30 (y-1) y^3)-3 m_c^3 s^2 x^2 (y-1)^2\\
  \nonumber& & y^2 (32 x^7 (y-1)^5-24 x^6 (y-34) (y-1)^4+x^5 (y-1)^3 (y (220 y+109)+1536)-x^4 (y-1)^2 (y (y (492 y+4531)\\
  \nonumber& &+410)-752)+x^3 (y-1) y (4 y (57 y^2+552 y-34)-543)+x^2 y^2 (y (9 y (4 y+183)-1418)-30)+x y^3 (y (32 y\\
  \nonumber& &-313)+211)-32 (y-1) y^4)+5 m_c s^3 x^3 (y-1)^3 y^3 (6 x^6 (y-1)^4+2 x^5 (y-1)^3 (4 y+81)+x^4 (y-1)^2 (y (51 y\\
  \nonumber& &+307)+306)-x^3 (y-1) (y (y (62 y+997)-306)-150)+x^2 y (y (241-5 y (3 y+56))+7)+2 x y^3 (3 y+4)\\
  \nonumber& &+6 (y-1) y^3))\},\\
  \nonumber\rho _{\langle \bar{q} Gq\rangle}^{0,\Lambda _cD^*}(s)&=&\langle g_s \bar{q}\sigma Gq\rangle \int_{x_{min}}^{x_{max}}  \,dx \int_{y_{min}}^{y_{max}}  \,dy \Bigg\{ \frac{-1}{24576 \pi^6 x^2 y^2} \{C_{11} (x-1) (x (y-1) (m_c^2-s y)-m_c^2 y) (C_{21} (m_c^4 (7 (x-1) y\\
  \nonumber& &-7 x+12) (x (-y)+x+y)^2-2 m_c^2 s x (y-1) y ((x-1) y-x) (43 (x-1) y-43 x+48)+s^2 x^2 (y-1)^2 y^2 (109 \\
  \nonumber& &(x-1) y-109 x+114))+C_{22} (m_c^4 (x (y-1)-y) (28 x^2 (y-1)^2-x (56 y-29) (y-1)+y (28 y-29)-18)\\
  \nonumber& &-2 m_c^2 s x (y-1) y (76 x^2 (y-1)^2-x (152 y-77) (y-1)+y (76 y-77)-9)+s^2 x^2 (y-1)^2 y^2 (160 x (y-1)\\
  \nonumber& &-160 y+161)))\}+\frac{1}{1769472 \pi^6 x^3 y^3} \{C_{11} (x-1) (x (y-1) (m_c^2-s y)-m_c^2 y) (6 C_{21} (m_c^4 (x (-y)+x+y)^2 \\
  \nonumber& &(20 x^3 (y-1)^2+16 x^2 (y-1) (y+3)+x (8 (6-5 y) y+25)+4 y (y+1)+3)-m_c^2 s x (y-1) y (x (y-1)-y) \\
  \nonumber& &(199 x^3 (y-1)^2+x^2 (y-1) (155 y+411)+x ((222-383 y) y+209)+y (29 y-16)+3)+s^2 x^2 (y-1)^2 y^2 \\
  \nonumber& &(243 x^3 (y-1)^2+x^2 (y-1) (187 y+491)+x ((230-463 y) y+248)+y (33 y-28)))+C_{22} (m_c^4 (269 x^5 (y\\
  \nonumber& &-1)^4-2 x^4 (y-1)^3 (83 y-278)-x^3 (y-1)^2 (2 y (575 y+323)-287)+2 x^2 (y-1) y (y (878 y-219)-279)\\
  \nonumber& &+x y (y ((590-743 y) y+237)+6)+2 y^2 (y (17 y-14)+3))-2 m_c^2 s x (y-1) y (800 x^4 (y-1)^3+2 x^3 (y-1)^2\\
  \nonumber& & (176 y+809)-x^2 (y-1) (3162 y^2+351 y-818)+x y (2 y (1034 y-687)-775)+y ((49-58 y) y-3))+s^2\\
  \nonumber& & x^2 (y-1)^2 y^2 (1763 x^3 (y-1)^2+4 x^2 (y-1) (591 y+886)+x ((2500-4209 y) y+1781)+2 y (41 y-35))))\}\Bigg\} ,\\
  \nonumber\rho _{\langle \bar{q} q\rangle^2}^{0,\Lambda _cD^*}(s)&=&\langle \bar{q} q\rangle^2 \int_{x_{min}}^{x_{max}}  \,dx \int_{y_{min}}^{y_{max}}  \,dy  \frac{-1}{1536 \pi^4 x^2 y^2} \{C_{11} (x-1) (x (y-1) (m_c^2-s y)-m_c^2 y) (m_c^3 y^2 (20 C_{21}+23 C_{22})\\
  \nonumber& &+2 m_c x^2 (y-1)^2 (2 m_c^2 (5 C_{21}+3 C_{22})-s y (31 C_{21}+18 C_{22}))+m_c x (y-1) y (s y (38 C_{21}+47 C_{22})\\
  \nonumber& &-5 m_c^2 (8 C_{21}+7 C_{22}))) \},\\
  \nonumber\rho _{\langle \bar{q} q\rangle\langle \bar{q}G q\rangle}^{0,\Lambda _cD^*}(s)&=&\langle \bar{q} q\rangle \langle g_s^2 \bar{q}\sigma G q\rangle \int_{x_{min}}^{x_{max}}  \,dx \int_{y_{min}}^{y_{max}}  \,dy\Bigg\{  \frac{1}{1536 \pi^4 x y} \{C_{11} (m_c^3 y^2 (20 C_{21}+23 C_{22})+m_c x^2 (y-1)^2 (4 m_c^2 (5 C_{21}\\
  \nonumber& &+3 C_{22})-s y (41 C_{21}+24 C_{22}))+m_c x (y-1) y (s y (29 C_{21}+35 C_{22})-5 m_c^2 (8 C_{21}+7 C_{22})))\}\\
  \nonumber& &+\frac{-1}{36864 \pi^4 x^2 y^2} \{C_{11} m_c (4 C_{21} m_c^2 (x (y-1)-y) (30 x^3 (y-1)^2-5 x^2 (y-1) (11 y-6)-x y (17 y+24)+6 y^2)\\
  \nonumber& &-2 C_{21} s x (y-1) y (123 x^3 (y-1)^2-3 x^2 (y-1) (68 y-41)-4 x y (12 y+19)+21 y^2)+C_{22} (s x (y-1) y (2 x^3\\
  \nonumber& & (y-1)^2+2 x^2 (y-1) (99 y+1)+x y (23 y+201)-12 y^2)-m_c^2 (x (y-1)-y) (4 x^3 (y-1)^2+4 x^2 (y-1) (32 y\\
  \nonumber& &+1)+19 x y (y+7)-12 y^2))) \}\Bigg\},\\
  \nonumber\rho _{\langle \bar{q}G q\rangle^2}^{0,\Lambda _cD^*}(s)&=&\langle g_s^2 \bar{q}\sigma G q\rangle ^2 \int_{x_{min}}^{x_{max}}  \,dx \int_{y_{min}}^{y_{max}}  \,dy  \Bigg\{\frac{-1}{1769472 \pi^4 x y} \{C_{11} m_c (24 C_{21} (5 x^2 (y-1)^2-x (y-3) (y-1)+5 y^2)+C_{22} \\
  \nonumber& &(50 x^2 (y-1)^2+x (173 y+16) (y-1)+4 y (2 y+1)))\}+\frac{1}{147456 \pi^4 x y} \{C_{11} m_c (4 x^2 (y-1)^2 (30 C_{21}-C_{22})-x y\\
  \nonumber& & (y-1) (96 C_{21}+133 C_{22})-12 y^2 (2 C_{21}+C_{22})) \}\Bigg\}+\langle g_s^2 \bar{q}\sigma G q\rangle ^2 \int_{0}^{1}  \,dx \int_{0}^{1}  \,dy\Bigg\{ \frac{1}{442368 \pi^4 x^2 (y-1) y^2} \{C_{11} \\
  \nonumber& &m_c^3 (x (y-1)-y) (3 C_{21} (x^2 (y-1)^2+x ((7-9 y) y+2)+y (y+5))+C_{22} (x (y-1) (x (y-1)+25 y)+6 y))\} \\
  \nonumber& &+\frac{-1}{73728 \pi^4 x^2 (y-1) y^2} \{C_{11} m_c^3 (x (y-1)-y) (x^2 (y-1)^2 (63 C_{21}+C_{22})-2 x y (y-1) (14 C_{21}+17 C_{22})\\
  \nonumber& &-9 C_{21} y^2)\}\Bigg\}\delta (s-\mathcal{F} (m_c^2)),\\
\rho _{\langle \bar{q} q\rangle^3}^{0,\Lambda _cD^*}(s)&=&\langle \bar{q} q\rangle ^3 \int_{x_{min}}^{x_{max}}  \,dx   \frac{1}{288 \pi^2} \{C_{11} (13 C_{21} m_c^2+3 C_{21} s (x-1) x+7 C_{22} m_c^2+C_{22} s (x-1) x) \},
\end{eqnarray}
}
where the parameters $C_{11}=C_{21}=C_{12}=C_{22}=1$ for the current $J^{\Lambda _cD^*}(s)$ and $C_{11}=C_{21}=1$, $C_{12}=C_{22}=-1$ for the current $J ^{\varSigma _cD^*}(s)$. 
The spectral densities for the interpolating currents $J^{\Lambda  _c^*D}$, $J^{\Lambda  _c^*D^*}$, $J^{\Sigma _c^*D}$ and $J^{\Sigma _c^*D^*}$ are obtained as
{\allowdisplaybreaks
\begin{eqnarray}
  \nonumber\rho _{pert}^{1,\Lambda  _c^*D}(s)&=&\int_{x_{min}}^{x_{max}}  \,dx \int_{y_{min}}^{y_{max}}  \,dy \frac{1}{39321600 \pi^8 x^4 y^4}\{ (x-1)^4 (y-1) (m_c^2 (x (-y)+x+y)+s x (y-1) y)^3 (m_c^4 (x (y-1)-y) \\
  \nonumber& &(259 x^2 (y-1)^2-2 x (259 y-247) (y-1)+y (259 y-494)-110)-m_c^2 s x (y-1) y (1843 x^2 (y-1)^2-x \\
  \nonumber& &(3686 y-2673) (y-1)+y (1843 y-2673)-110)+s^2 x^2 (y-1)^2 y^2 (2384 x (y-1)-2384 y+2979))\},\\
  \nonumber\rho _{\langle \bar{q}q\rangle}^{1,\Lambda  _c^*D}(s)&=&\langle \bar{q}q\rangle \int_{x_{min}}^{x_{max}}  \,dx \int_{y_{min}}^{y_{max}}  \,dy \frac{1}{73728 \pi^6 x^3 y^3} \{m_c (x-1)^2 (m_c^2 (x (-y)+x+y)+s x (y-1) y)^2 (m_c^2 (x (y-1)-y) \\
  \nonumber& & (125 x^2 (y-1)^2-2 x (95 y-61) (y-1)+y (65 y-98))-s x (y-1) y (245 x^2 (y-1)^2-2 x (215 y\\
  \nonumber& &-121) (y-1)+y (185 y-218)))\},\\
  \nonumber\rho _{\langle GG\rangle}^{1,\Lambda  _c^*D}(s)&=&\langle g_s^2GG\rangle \int_{x_{min}}^{x_{max}}  \,dx \int_{y_{min}}^{y_{max}}  \,dy \frac{-1}{566231040 \pi^8 x^4 y^4}\{ (x-1)^2 (y-1) (m_c^6 (x (-y)+x+y)^2 (3108 x^6 (y-1)^4-\\
  \nonumber& &3 x^5 (y-1)^3 (6155 y-4068)+2 x^4 (y-1)^2 (y (9624 y-20407)+6882)+x^3 (y-1) (y (2 y (9949 y-2299)\\
  \nonumber& &-22301)+3348)-4 x^2 (y (y (y (9028 y-16947)+4114)+120)+330)+x y (y (y (9215 y-13161)-132)\\
  \nonumber& &+528)+12 y^2 (y (259 y-499)-110))-6 m_c^4 s x (y-1) y (x (y-1)-y) (2096 x^6 (y-1)^4-8 x^5 (y-1)^3\\
  \nonumber& & (1715 y-942)+x^4 (y-1)^2 (y (11756 y-28829)+8432)+x^3 (y-1) (y (25436 y^2-9238 y-14901)\\
  \nonumber& &+2640)-2 x^2 (y (y (17 y (1166 y-1937)+8661)-60)+176)+x y (y (y (11980 y-15811)+148)+88)\\
  \nonumber& &+16 y^2 (y (131 y-209)-22))+3 m_c^2 s^2 x^2 (y-1)^2 y^2 (3476 x^6 (y-1)^4-3 x^5 (y-1)^3 (10797 y-3988)+\\
  \nonumber& &2 x^4 (y-1)^2 (y (10900 y-32619)+6618)+x^3 (y-1) (y (96070 y^2-43066 y-31823)+4484)-4 x^2 (y (y\\
  \nonumber& & (y (35470 y-54949)+14264)-256)+66)+x y^2 (49449 y^2-62043 y+964)+4 y^2 (y (869 y-1253)\\
  \nonumber& &-66))+2 s^3 x^4 (y-1)^3 y^4 (15219 x^3 (y-1)^2+11 x^2 (y-1) (873 y+2627)+x (19 (3659-3637 y) y+12742)\\
  \nonumber& &+y (44281 y-53652)+936))\},\\
  \nonumber\rho _{\langle \bar{q} Gq\rangle}^{1,\Lambda  _c^*D}(s)&=&\langle g_s \bar{q}\sigma Gq\rangle \int_{x_{min}}^{x_{max}}  \,dx \int_{y_{min}}^{y_{max}}  \,dy \Bigg\{ \frac{-1}{1179648 \pi^6 x^3 y^3}\{ (x-1) (m_c^5 (x (-y)+x+y)^2 (156 x^4 (y-1)^3-2 x^3 (y-1)^2\\
  \nonumber& & (289 y-201)-x^2 (y-1) (y (1145 y+877)-246)+x y (5 y (683 y-802)-221)+264 (10-7 y) y^2)-2 m_c^3 \\
  \nonumber& &s x (y-1) y (x (y-1)-y) (248 x^4 (y-1)^3-2 x^3 (y-1)^2 (445 y-293)-x^2 (y-1) (y (1853 y+1269)-338)\\
  \nonumber& &+x y (y (5399 y-5914)-301)+264 (14-11 y) y^2)+m_c s^2 x^2 (y-1)^2 y^2 (340 x^4 (y-1)^3-2 x^3 (y-1)^2 (601 y\\
  \nonumber& &-385)-x^2 (y-1) (y (2561 y+1661)-430)+3 x y (y (2461 y-2606)-127)-792 y^2 (5 y-6)))\}\\
  \nonumber& &+\frac{-1}{32768 \pi^6 x^2 y^2}\{ (x-1) (m_c^5 (x (-y)+x+y)^2 (125 x^2 (y-1)^2-x (190 y-123) (y-1)+y (65 y-87))-2 m_c^3\\
  \nonumber& & s x (y-1) y (x (y-1)-y) (165 x^2 (y-1)^2-x (270 y-163) (y-1)+y (105 y-127))+m_c s^2 x^2 (y-1)^2 y^2 (205\\
  \nonumber& & x^2 (y-1)^2-7 x (50 y-29) (y-1)+y (145 y-167)))\}\Bigg\} ,\\
  \nonumber\rho _{\langle \bar{q} q\rangle^2}^{1,\Lambda  _c^*D}(s)&=&\langle \bar{q} q\rangle^2 \int_{x_{min}}^{x_{max}}  \,dx \int_{y_{min}}^{y_{max}}  \,dy  \frac{1}{3072 \pi^4 x y}\{ (x-1) (y-1) (m_c^2 (-23 x (y-1)+23 y-86)+37 s x (y-1) y) (m_c^2 (x (\\
  \nonumber& &-y)+x+y)+s x (y-1) y) \},\\
  \nonumber\rho _{\langle \bar{q} q\rangle\langle \bar{q}G q\rangle}^{1,\Lambda  _c^*D}(s)&=&\langle \bar{q} q\rangle \langle g_s^2 \bar{q}\sigma G q\rangle \int_{x_{min}}^{x_{max}}  \,dx \int_{y_{min}}^{y_{max}}  \,dy\Bigg\{  \frac{1}{3072 \pi^4}\{ (y-1) (m_c^2 (-23 x (y-1)+23 y-43)+30 s x (y-1) y)\}\\
  \nonumber& &+\frac{1}{73728 \pi^4 x y}\{ (y-1) (m_c^2 (-12 x^3 (y-1)^2+x^2 (y-1) (123 y-110)-x (y (227 y+832)+98)+4 y (29 y+263))\\
  \nonumber& &+s x (y-1) y (2 x (9 x (y-1)-80 y+9)+157 y))\}\Bigg\},\\
  \nonumber\rho _{\langle \bar{q}G q\rangle^2}^{1,\Lambda  _c^*D}(s)&=&\langle g_s^2 \bar{q}\sigma G q\rangle ^2 \int_{x_{min}}^{x_{max}}  \,dx \int_{y_{min}}^{y_{max}}  \,dy  \Bigg\{\frac{-1}{3538944 \pi^4}\{ (y-1) (295 x (y-1)-339 y+277)\}+\frac{1}{73728 \pi^4}\{ (y-1) (3 x (y-1)\\
  \nonumber& &-29 y)\}+\langle g_s^2 \bar{q}\sigma G q\rangle ^2 \int_{0}^{1}  \,dx \int_{0}^{1}  \,dy\Bigg\{ \frac{1}{3538944 \pi^4 x y}\{ m_c^2 (193 x^2 (y-1)^2-x (402 y-223) (y-1)+y (209 y-\\
  \nonumber& &227)+8) \}+\frac{1}{294912 \pi^4 x y}\{ m_c^2 (-6 x^2 (y-1)^2+x (47 y+98) (y-1)+(1052-41 y) y)\}\Bigg\}\delta (s-\mathcal{F} (m_c^2))\},\\
 \rho _{\langle \bar{q} q\rangle^3}^{1,\Lambda  _c^*D}(s)&=&\langle \bar{q} q\rangle ^3 \int_{x_{min}}^{x_{max}}  \,dx   \frac{1}{1152 \pi^2}\{ m_c-12 m_c x \},
\end{eqnarray}
}

{\allowdisplaybreaks
\begin{eqnarray}
  \nonumber\rho _{pert}^{0,\Lambda  _c^*D}(s)&=&\int_{x_{min}}^{x_{max}}  \,dx \int_{y_{min}}^{y_{max}}  \,dy \frac{1}{39321600 \pi^8 x^5 y^5}\{ (x-1)^4 (m_c^2 (x (y-1)-y)-s x (y-1) y)^4 (m_c^3 (x (y-1)-y) (38 x^2 (y-\\
  \nonumber& &1)^2-x (36 y+77) (y-1)-y (2 y+13))-m_c s x (y-1) y (103 x^2 (y-1)^2-x (96 y+287) (y-1)-y (7 y+33)))\},\\
  \nonumber\rho _{\langle \bar{q}q\rangle}^{0,\Lambda  _c^*D}(s)&=&\langle \bar{q}q\rangle \int_{x_{min}}^{x_{max}}  \,dx \int_{y_{min}}^{y_{max}}  \,dy \frac{-1}{147456 \pi^6 x^3 y^3} \{(x-1)^2 (m_c^2 (x (-y)+x+y)+s x (y-1) y)^2 (m_c^4 (x (y-1)-y) (21 x^2 \\
  \nonumber& &(y-1)^2-2 x (21 y-8) (y-1)+y (21 y-16)+28)-2 m_c^2 s x (y-1) y (196 x^2 (y-1)^2-x (392 y-207) (y-1)\\
  \nonumber& &+y (196 y-207)+14)+s^2 x^2 (y-1)^2 y^2 (611 x (y-1)-611 y+638))\},\\
  \nonumber\rho _{\langle GG\rangle}^{0,\Lambda  _c^*D}(s)&=&\langle g_s^2GG\rangle \int_{x_{min}}^{x_{max}}  \,dx \int_{y_{min}}^{y_{max}}  \,dy \frac{1}{283115520 \pi^8 x^5 y^5}\{ (x-1)^2 (m_c^7 (-(x (-y)+x+y)^3) (456 x^6 (y-1)^4-4 x^5 (y-1)^3\\
  \nonumber& & (221 y-12)+2 x^4 (y-1)^2 (y (55 y-122)-636)+2 x^3 (y-1) (y (y (519 y+1423)+52)-432)-x^2 y (y (y (494 y\\
  \nonumber& &+155)-1545)+536)-2 x y^2 (y (101 y+122)-208)-24 y^3 (y+4))-3 m_c^5 s x (y-1) y (x (-y)+x+y)^2 (510 \\
  \nonumber& &x^6 (y-1)^4-6 x^5 (y-1)^3 (163 y+20)+2 x^4 (y-1)^2 (y (23 y-87)-885)+x^3 (y-1) (y (y (1465 y+4414)+\\
  \nonumber& &216)-1140)-x^2 y (y (11 y (76 y+51)-2505)+588)+x y^2 (408-y (177 y+361))-30 y^3 (y+4))+3 m_c^3 s^2 x^2\\
  \nonumber& & (y-1)^2 y^2 (512 x^7 (y-1)^5-28 x^6 (y-1)^4 (53 y+8)+2 x^5 (y-1)^3 (y (455 y+32)-992)+2 x^4 (y-1)^2 (y (y\\
  \nonumber& & (1001 y+3067)+1170)-624)-x^3 (y-1) y (y (y (3218 y+7005)-3109)-792)+x^2 y^2 (y (y (1170 y+761)\\
  \nonumber& &-2871)+240)+2 x y^3 (y+4) (38 y-11)+32 y^4 (y+4))-m_c s^3 x^3 (y-1)^3 y^3 (462 x^6 (y-1)^4-2 x^5 (y-1)^3 (433 y\\
  \nonumber& &+132)-2 x^4 (y-1)^2 (y (107 y+101)+957)+x^3 (y-1) (y (y (2463 y+7526)+524)-1188)-5 x^2 y (y (y (364 y\\
  \nonumber& &+313)-885)+28)+5 x y^2 ((y-59) y-32)-30 y^3 (y+4)))\},\\
  \nonumber\rho _{\langle \bar{q} Gq\rangle}^{0,\Lambda  _c^*D}(s)&=&\langle g_s \bar{q}\sigma Gq\rangle \int_{x_{min}}^{x_{max}}  \,dx \int_{y_{min}}^{y_{max}}  \,dy \Bigg\{ \frac{-1}{3538944 \pi^6 x^3 y^3}\{ (x-1) (m_c^6 (-(x (-y)+x+y)^2) (34 x^4 (y-1)^3-8 x^3 (y-1)^2 (49 y\\
  \nonumber& &-4)+x^2 (y-1) (y (663 y-1238)+178)+x (y ((2029-286 y) y+648)+180)-y (y (19 y+842)+1560))+3 m_c^4 \\
  \nonumber& & s x (y-1) y (32 x^5 (y-1)^4-x^4 (y-1)^3 (487 y-97)+x^3 (y-1)^2 (4 y (308 y-323)+185)-2 x^2 (y-1) (y (y (547 y-\\
  \nonumber& &1428)-71)-60)+x y (y (8 y (35 y-278)-767)+920)+y^2 (y (37 y+600)+1040))-3 m_c^2 s^2 x^2 (y-1)^2 y^2 (26 x^4\\
  \nonumber& & (y-1)^3-22 x^3 (y-1)^2 (23 y-7)+x^2 (y-1) (y (879 y-1136)+188)+x (y ((1285-344 y) y+10)+60)-y (y (\\
  \nonumber& &55 y+358)+520))+s^3 x^3 (y-1)^3 y^3 (16 x^3 (y-1)^2-x^2 (y-1) (529 y-203)+11 x (40 y^2-78 y+17)+y (73 y+\\
  \nonumber& &116)))\}+\frac{1}{32768 \pi^6 x^2 y^2}\{ (x-1) (m_c^6 (x (-y)+x+y)^2 (14 x^2 (y-1)^2-x (28 y-11) (y-1)+y (14 y-11)+7)-\\
  \nonumber& &m_c^4 s x (y-1) y (x (y-1)-y) (217 x^2 (y-1)^2-x (434 y-221) (y-1)+y (217 y-221)+14)+m_c^2 s^2 x^2 (y-1)^2\\
  \nonumber& & y^2 (472 x^2 (y-1)^2-x (944 y-489) (y-1)+y (472 y-489)+7)-s^3 x^3 (y-1)^3 y^3 (269 x (y-1)-269 y+279))\}\Bigg\} ,\\
  \nonumber\rho _{\langle \bar{q} q\rangle^2}^{0,\Lambda  _c^*D}(s)&=&\langle \bar{q} q\rangle^2 \int_{x_{min}}^{x_{max}}  \,dx \int_{y_{min}}^{y_{max}}  \,dy  \frac{1}{512 \pi^4 x^2 y^2}\{ (x-1) (2 m_c^5 (9 x (y-1)-y) (x (-y)+x+y)^2-m_c^3 s x (y-1) y (53 x (y-1)\\
  \nonumber& &-17 y) (x (y-1)-y)+5 m_c s^2 x^2 (y-1)^2 y^2 (7 x (y-1)-3 y)) \},\\
  \nonumber\rho _{\langle \bar{q} q\rangle\langle \bar{q}G q\rangle}^{0,\Lambda  _c^*D}(s)&=&\langle \bar{q} q\rangle \langle g_s^2 \bar{q}\sigma G q\rangle \int_{x_{min}}^{x_{max}}  \,dx \int_{y_{min}}^{y_{max}}  \,dy\Bigg\{  \frac{1}{1024 \pi^4 x y}\{ m_c s x (y-1) y (53 x (y-1)-17 y)-4 m_c^3 (x (y-1)-y) (9 x (y-1)\\
  \nonumber& &-y)\}+\frac{1}{73728 \pi^4 x^2 y^2} 2 m_c^3 (x (y-1)-y) (68 x^3 (y-1)^2-x^2 (y-1) (33 y-68)+x y (203 y+31)-261 y^2)-m_c s x \\
  \nonumber& &(y-1) y (132 x^3 (y-1)^2+x^2 (y-1) (35 y+132)+x y (1363 y+144)-1574 y^2)\}\Bigg\},\\
  \nonumber\rho _{\langle \bar{q}G q\rangle^2}^{0,\Lambda  _c^*D}(s)&=&\langle g_s^2 \bar{q}\sigma G q\rangle ^2 \int_{x_{min}}^{x_{max}}  \,dx \int_{y_{min}}^{y_{max}}  \,dy  \Bigg\{\frac{1}{3538944 \pi^4 x y}\{ m_c (-120 x^2 (y-1)^2+x ((5-61 y) y+56)+2 y (668 y-107))\}\\
  \nonumber& &+\frac{-1}{147456 \pi^4 x y}\{ m_c (68 x^2 (y-1)^2+31 x y (y-1)+261 y^2)\}+\langle g_s^2 \bar{q}\sigma G q\rangle ^2 \int_{0}^{1}  \,dx \int_{0}^{1}  \,dy\Bigg\{ \frac{1}{1769472 \pi^4 x^2 (y-1) y^2}\\
  \nonumber& & \{m_c^3 (x (y-1)-y) (4 x^2 (y-1)^2-x (16 y-9) (y-1)-2 y (y+5)) \}+\frac{-1}{147456 \pi^4 x^2 (y-1) y^2}\{ m_c^3 (x (y-1)-y) \\
  \nonumber& &(2 x^2 (y-1)^2-41 x y (y-1)-526 y^2)\}\Bigg\}\delta (s-\mathcal{F} (m_c^2))\},\\
  \rho _{\langle \bar{q} q\rangle^3}^{0,\Lambda  _c^*D}(s)&=&\langle \bar{q} q\rangle ^3 \int_{x_{min}}^{x_{max}}  \,dx   \frac{-1}{1152 \pi^2} 5 (13 m_c^2+3 s (x-1) x) \},
\end{eqnarray}
}

{\allowdisplaybreaks
\begin{eqnarray}
  \nonumber\rho _{pert}^{1,\Lambda  _c^*D^*}(s)&=&\int_{x_{min}}^{x_{max}}  \,dx \int_{y_{min}}^{y_{max}}  \,dy \frac{1}{9830400 \pi^8 x^4 y^4}\{ (x-1)^4 (y-1) (m_c^2 (x (-y)+x+y)+s x (y-1) y)^3 (m_c^4 (x (y-1)\\
  \nonumber& &-y) (77 x^2 (y-1)^2-x (154 y-197) (y-1)+y (77 y-197)-80)-2 m_c^2 s x (y-1) y (292 x^2 (y-1)^2\\
  \nonumber& &-x (584 y-527) (y-1)+y (292 y-527)-40)+s^2 x^2 (y-1)^2 y^2 (827 x (y-1)-827 y+1177))\},\\
  \nonumber\rho _{\langle \bar{q}q\rangle}^{1,\Lambda  _c^*D^*}(s)&=&\langle \bar{q}q\rangle \int_{x_{min}}^{x_{max}}  \,dx \int_{y_{min}}^{y_{max}}  \,dy \frac{1}{9216 \pi^6 x^3 y^3}\{ m_c (x-1)^2 (m_c^2 (x (-y)+x+y)+s x (y-1) y)^2 (m_c^2 (x (y-1)-y) \\
  \nonumber& &(16 x^2 (y-1)^2-x (29 y-16) (y-1)+y (13 y-22))-s x (y-1) y (40 x^2 (y-1)^2-x (77 y-40) (y-1)\\
  \nonumber& &+y (37 y-46)))\},\\
  \nonumber\rho _{\langle GG\rangle}^{1,\Lambda  _c^*D^*}(s)&=&\langle g_s^2GG\rangle \int_{x_{min}}^{x_{max}}  \,dx \int_{y_{min}}^{y_{max}}  \,dy \frac{-1}{283115520 \pi^8 x^4 y^4}\{ (x-1)^2 (y-1) (m_c^6 (x (-y)+x+y)^2 (1848 x^6 (y-1)^4-8 x^5\\
  \nonumber& & (y-1)^3 (694 y-1053)+x^4 (y-1)^2 (y (3611 y-19060)+9384)-x^3 (y-1) (y (y (3229 y-18970)+8452)\\
  \nonumber& &-888)+x^2 (y (y (y (10441 y-30180)+2091)+4288)-1920)+x y (y (3 (7101-2989 y) y+5056)+768)\\
  \nonumber& &+24 y^2 (y (77 y-197)-80))-3 m_c^4 s x (y-1) y (x (y-1)-y) (2608 x^6 (y-1)^4-4 x^5 (y-1)^3 (1899 y-2620)\\
  \nonumber& &+x^4 (y-1)^2 (y (3967 y-22828)+12112)-x^3 (y-1) (y (y (5033 y-23714)+11824)-3216)+x^2 (y (y (y \\
  \nonumber& &(18117 y-43444)+9519)+3152)-1024)+x y (y ((27887-14671 y) y+3408)+256)+16 y^2 (y (163 y-\\
  \nonumber& &329)-64))+3 m_c^2 s^2 x^2 (y-1)^2 y^2 (2248 x^6 (y-1)^4-8 x^5 (y-1)^3 (769 y-1055)+x^4 (y-1)^2 (y (131 y-\\
  \nonumber& &18020)+9752)-x^3 (y-1) (y (9 y (461 y-2650)+9724)-3176)+x^2 (y (y (y (23521 y-54020)+17139)+\\
  \nonumber& &2144)-384)+x y^2 ((30343-17847 y) y+2144)+8 y^2 (y (281 y-493)-48))-s^3 x^4 (y-1)^3 y^4 (1084 x^3 (y\\
  \nonumber& &-1)^2-x^2 (y-1) (9117 y+28)+2 x ((8523-4079 y) y+76)+3 y (5397 y-8789)-1264))\},\\
  \nonumber\rho _{\langle \bar{q} Gq\rangle}^{1,\Lambda  _c^*D^*}(s)&=&\langle g_s \bar{q}\sigma Gq\rangle \int_{x_{min}}^{x_{max}}  \,dx \int_{y_{min}}^{y_{max}}  \,dy \Bigg\{ \frac{-1}{3072 \pi^6 x^2 y^2}\{ m_c (x-1)^2 (y-1) (2 x (y-1)-2 y+1) (m_c^2 (x (-y)+x+y)\\
  \nonumber& &+s x (y-1) y) (m_c^2 (x (-y)+x+y)+2 s x (y-1) y)\}+\frac{-1}{4096 \pi^6 x^2 y^2}\{ (x-1) (m_c^5 (x (-y)+x+y)^2 (16 x^2 (y-1)^2\\
  \nonumber& &-x (29 y-16) (y-1)+y (13 y-19))-6 m_c^3 s x (y-1) y (x (y-1)-y) (8 x^2 (y-1)^2-x (15 y-8) (y-1)\\
  \nonumber& &+y (7 y-9))+m_c s^2 x^2 (y-1)^2 y^2 (32 x^2 (y-1)^2-x (61 y-32) (y-1)+y (29 y-35)))\}\Bigg\} ,\\
  \nonumber\rho _{\langle \bar{q} q\rangle^2}^{1,\Lambda  _c^*D^*}(s)&=&\langle \bar{q} q\rangle^2 \int_{x_{min}}^{x_{max}}  \,dx \int_{y_{min}}^{y_{max}}  \,dy  \frac{1}{192 \pi^4 x y}\{ (x-1) (y-1) (m_c^2 (x (-y)+x+y-10)+2 s x (y-1) y) (m_c^2 (x (-y)\\
  \nonumber& &+x+y)+s x (y-1) y) \},\\
  \nonumber\rho _{\langle \bar{q} q\rangle\langle \bar{q}G q\rangle}^{1,\Lambda  _c^*D^*}(s)&=&\langle \bar{q} q\rangle \langle g_s^2 \bar{q}\sigma G q\rangle \int_{x_{min}}^{x_{max}}  \,dx \int_{y_{min}}^{y_{max}}  \,dy\Bigg\{  \frac{1}{384 \pi^4}\{ (y-1) (2 m_c^2 (x (-y)+x+y-5)+3 s x (y-1) y)\}+\frac{-1}{3072 \pi^4 x y}\{ (x\\
  \nonumber& &-1) (y-1) (7 x (y-1)-4 y) (2 m_c^2 (x (-y)+x+y)+3 s x (y-1) y)\}\Bigg\},\\
  \nonumber\rho _{\langle \bar{q}G q\rangle^2}^{1,\Lambda  _c^*D^*}(s)&=&\langle g_s^2 \bar{q}\sigma G q\rangle ^2 \int_{x_{min}}^{x_{max}}  \,dx \int_{y_{min}}^{y_{max}}  \,dy  \Bigg\{\frac{1}{442368 \pi^4}\{ (y-1) (62 x (y-1)-75 y+65)\}+\frac{-1}{6144 \pi^4}\{ (y-1) (7 x (y-1)-4 y)\}\\
  \nonumber& &+\langle g_s^2 \bar{q}\sigma G q\rangle ^2 \int_{0}^{1}  \,dx \int_{0}^{1}  \,dy\Bigg\{ \frac{1}{221184 \pi^4 x y}\{ m_c^2 (-16 x^2 (y-1)^2+7 x (5 y-4) (y-1)+(28-19 y) y-12) \}\\
  \nonumber& &+\frac{1}{12288 \pi^4 x y}\{ m_c^2 (7 x (y-1)-4 y) (x (y-1)-y)\}\Bigg\}\delta (s-\mathcal{F} (m_c^2))\},\\
  \rho _{\langle \bar{q} q\rangle^3}^{1,\Lambda  _c^*D^*}(s)&=&\langle \bar{q} q\rangle ^3 \int_{x_{min}}^{x_{max}}  \,dx   \frac{-1}{48 \pi^2}\{ m_c x \},
\end{eqnarray}
}

{\allowdisplaybreaks
\begin{eqnarray}
  \nonumber\rho _{pert}^{0,\Lambda  _c^*D^*}(s)&=&\int_{x_{min}}^{x_{max}}  \,dx \int_{y_{min}}^{y_{max}}  \,dy \frac{1}{9830400 \pi^8 x^4 y^5}\{ m_c (x-1)^4 (y-1) (m_c^2 (x (y-1)-y)-s x (y-1) y)^4 (m_c^2 (7 x (y-1)-7 y\\
  \nonumber& &-23) (x (y-1)-y)-s x (y-1) y (37 x (y-1)-37 y-143))\},\\
  \nonumber\rho _{\langle \bar{q}q\rangle}^{0,\Lambda  _c^*D^*}(s)&=&\langle \bar{q}q\rangle \int_{x_{min}}^{x_{max}}  \,dx \int_{y_{min}}^{y_{max}}  \,dy \frac{-1}{73728 \pi^6 x^3 y^3}\{ (x-1)^2 (m_c^2 (x (-y)+x+y)+s x (y-1) y)^2 (m_c^4 (x (y-1)-y) (5 x^2\\
  \nonumber& & (y-1)^2-2 x (5 y-3) (y-1)+y (5 y-6)+16)-2 m_c^2 s x (y-1) y (101 x^2 (y-1)^2-2 x (101 y-65) (y-1)\\
  \nonumber& &+y (101 y-130)+8)+s^2 x^2 (y-1)^2 y^2 (389 x (y-1)-389 y+446))\},\\
  \nonumber\rho _{\langle GG\rangle}^{0,\Lambda  _c^*D^*}(s)&=&\langle g_s^2GG\rangle \int_{x_{min}}^{x_{max}}  \,dx \int_{y_{min}}^{y_{max}}  \,dy \frac{1}{141557760 \pi^8 x^4 y^5}\{ (x-1)^2 (m_c^7 (-(x (-y)+x+y)^3) (168 x^5 (y-1)^4-9 x^4 (y\\
  \nonumber& &-1)^3 (43 y+24)+x^3 (y-1)^2 (y (1155 y-152)-936)-x^2 (y-1) (y (6 y (278 y-183)+43)+552)+x y (y(y \\
  \nonumber& & (799 y+194)-795)-278)+y^2 (288-y (67 y+141)))-3 m_c^5 s x (y-1) y (x (-y)+x+y)^2 (260 x^5 (y-1)^4-\\
  \nonumber& &x^4 (y-1)^3 (583 y+420)+x^3 (y-1)^2 (y (1631 y-164)-1620)-x^2 (y-1) (y (2180 y^2-2034 y-41)+940)+\\
  \nonumber& &x y (y (y (959 y+494)-1155)-378)+y^2 (388-y (87 y+221)))+3 m_c^3 s^2 x^2 (y-1)^2 y^2 (304 x^6 (y-1)^5-3 x^5\\
  \nonumber& & (y-1)^4 (329 y+176)+2 x^4 (y-1)^3 (y (1371 y+128)-984)-x^3 (y-1)^2 (y (y (4703 y-3242)-2093)+1136\\
  \nonumber& &)+x^2 (y-1) y (y (y (3667 y-2272)-1640)+850)-2 x y^2 (y (y (541 y+398)-984)+5)+y^3 (y (59 y+157\\
  \nonumber& &)-296))-m_c s^3 x^3 (y-1)^3 y^3 (300 x^5 (y-1)^4-3 x^4 (y-1)^3 (229 y+180)+x^3 (y-1)^2 (y (2439 y-476)-1980\\
  \nonumber& &)-x^2 (y-1) (y (18 y (170 y-217)-209)+1140)+x y (y (y (991 y+806)-1875)-2)+y^2 (17 y (y+3)+12)))\},\\
  \nonumber\rho _{\langle \bar{q} Gq\rangle}^{0,\Lambda  _c^*D^*}(s)&=&\langle g_s \bar{q}\sigma Gq\rangle \int_{x_{min}}^{x_{max}}  \,dx \int_{y_{min}}^{y_{max}}  \,dy \Bigg\{ \frac{-1}{442368 \pi^6 x^2 y^3}\{ (x-1)^2 (y-1) (s x (y-1) y (16 x (y-1)-202 y+13)-m_c^2 (7 x \\
  \nonumber& &(y-1)-85 y+4) (x (y-1)-y)) (m_c^2 (x (-y)+x+y)+s x (y-1) y)^2\}+\frac{1}{24576 \pi^6 x^2 y^2}\{ (x-1) (m_c^6 (x (-y)+x\\
  \nonumber& &+y)^2 (5 x^2 (y-1)^2-5 x (2 y-1) (y-1)+5 y (y-1)+6)-3 m_c^4 s x (y-1) y (x (y-1)-y) (53 x^2 (y-1)^2-x (106 y\\
  \nonumber& &-63) (y-1)+y (53 y-63)+4)+3 m_c^2 s^2 x^2 (y-1)^2 y^2 (133 x^2 (y-1)^2-x (266 y-153) (y-1)+y (133 y-153)\\
  \nonumber& &+2)-5 s^3 x^3 (y-1)^3 y^3 (49 x (y-1)-49 y+55))\}\Bigg\} ,\\
  \nonumber\rho _{\langle \bar{q} q\rangle^2}^{0,\Lambda  _c^*D^*}(s)&=&\langle \bar{q} q\rangle^2 \int_{x_{min}}^{x_{max}}  \,dx \int_{y_{min}}^{y_{max}}  \,dy  \frac{1}{384 \pi^4 x^2 y^2}\{ (x-1) (m_c^5 (11 x (y-1)-2 y) (x (-y)+x+y)^2-4 m_c^3 s x (y-1) y \\
  \nonumber& &(11 x (y-1)-5 y) (x (y-1)-y)+3 m_c s^2 x^2 (y-1)^2 y^2 (11 x (y-1)-6 y)) \},\\
  \nonumber\rho _{\langle \bar{q} q\rangle\langle \bar{q}G q\rangle}^{0,\Lambda  _c^*D^*}(s)&=&\langle \bar{q} q\rangle \langle g_s^2 \bar{q}\sigma G q\rangle \int_{x_{min}}^{x_{max}}  \,dx \int_{y_{min}}^{y_{max}}  \,dy\Bigg\{  \frac{1}{384 \pi^4 x y} 2 m_c s x (y-1) y (11 x (y-1)-5 y)-m_c^3 (11 x (y-1)-2 y) (x (y-1)\\
  \nonumber& &-y)\}+\frac{1}{18432 \pi^4 x y^2}\{ m_c^3 (x (y-1)-y) (48 x^2 (y-1)^2-x (103 y-48) (y-1)+y (31 y-40))-m_c s x (y-1) y (96 x^2 (y\\
  \nonumber& &-1)^2-x (215 y-96) (y-1)+y (95 y-88))\}\Bigg\},\\
  \nonumber\rho _{\langle \bar{q}G q\rangle^2}^{0,\Lambda  _c^*D^*}(s)&=&\langle g_s^2 \bar{q}\sigma G q\rangle ^2 \int_{x_{min}}^{x_{max}}  \,dx \int_{y_{min}}^{y_{max}}  \,dy  \Bigg\{\frac{-1}{884736 \pi^4 x y}\{ m_c (8 x^2 (y-1)^2+30 x y (y-1)+(19-13 y) y)\}+\frac{-1}{9216 \pi^4 y}\{ m_c (y\\
  \nonumber& &-1) (6 x (y-1)-5 y)\}+\langle g_s^2 \bar{q}\sigma G q\rangle ^2 \int_{0}^{1}  \,dx \int_{0}^{1}  \,dy\Bigg\{ \frac{1}{73728 \pi^4 x y}\{ m_c^3 (x (y-1)-y)\}+\frac{1}{1536 \pi^4 x y^2}\{ m_c^3 (x (-y)\\
  \nonumber& &+x+y)^2\}\Bigg\}\delta (s-\mathcal{F} (m_c^2))\},\\
  \rho _{\langle \bar{q} q\rangle^3}^{0,\Lambda  _c^*D^*}(s)&=&\langle \bar{q} q\rangle ^3 \int_{x_{min}}^{x_{max}}  \,dx   \frac{-1}{288 \pi^2} 31 m_c^2+9 s (x-1) x \},
\end{eqnarray}
}

{\allowdisplaybreaks
\begin{eqnarray}
  \nonumber\rho _{pert}^{1,\varSigma _c^*D}(s)&=&\int_{x_{min}}^{x_{max}}  \,dx \int_{y_{min}}^{y_{max}}  \,dy \frac{1}{19660800 \pi^8 x^4 y^4}\{ (x-1)^4 (y-1) (m_c^2 (x (-y)+x+y)+s x (y-1) y)^3 (m_c^4 (x (y-1)-y) \\
  \nonumber& &(104 x^2 (y-1)^2-2 x (104 y-167) (y-1)+2 y (52 y-167)-45)-m_c^2 s x (y-1) y (838 x^2 (y-1)^2-x (1676 y\\
  \nonumber& &-1593) (y-1)+y (838 y-1593)-45)+21 s^2 x^2 (y-1)^2 y^2 (54 x (y-1)-54 y+79))\},\\
  \nonumber\rho _{\langle \bar{q}q\rangle}^{1,\varSigma _c^*D}(s)&=&\langle \bar{q}q\rangle \int_{x_{min}}^{x_{max}}  \,dx \int_{y_{min}}^{y_{max}}  \,dy \frac{1}{12288 \pi^6 x^3 y^3}\{ m_c (x-1)^2 (x (y-1)-y) (m_c^2 (x (y-1)-y)-s x (y-1) y)^2 (m_c^2 (21 x\\
  \nonumber& & (y-1)-9 y+20) (x (y-1)-y)-s x (y-1) y (41 x (y-1)-29 y+40))\},\\
  \nonumber\rho _{\langle GG\rangle}^{1,\varSigma _c^*D}(s)&=&\langle g_s^2GG\rangle \int_{x_{min}}^{x_{max}}  \,dx \int_{y_{min}}^{y_{max}}  \,dy \frac{-1}{283115520 \pi^8 x^4 y^4}\{ (x-1)^2 (y-1) (m_c^6 (x (-y)+x+y)^2 (1248 x^6 (y-1)^4-x^5 (y\\
  \nonumber& &-1)^3 (8801 y-6444)+x^4 (y-1)^2 (y (11177 y-22920)+8604)+x^3 (y-1) (y (y (5707 y-2390)-13783)\\
  \nonumber& &+2868)+x^2 (y (y ((37760-14733 y) y-9897)+120)-540)+2 x y (y (y (2077 y-4148)+258)+108)+\\
  \nonumber& &12 y^2 (y (104 y-329)-45))-6 m_c^4 s x (y-1) y (x (y-1)-y) (920 x^6 (y-1)^4-x^5 (y-1)^3 (6775 y-3896)+\\
  \nonumber& &x^4 (y-1)^2 (y (7065 y-16064)+4888)+x^3 (y-1) (y (y (9875 y-4433)-9079)+1768)+x^2 (y (y ((35328\\
  \nonumber& &-17525 y) y-9498)+174)-144)+x y (y (y (5520 y-9311)+270)+36)+8 y^2 (y (115 y-257)-18))+\\
  \nonumber& &3 m_c^2 s^2 x^2 (y-1)^2 y^2 (1584 x^6 (y-1)^4-x^5 (y-1)^3 (16443 y-6124)+x^4 (y-1)^2 (y (14251 y-36648)+7388)\\
  \nonumber& &+x^3 (y-1) (y (y (40801 y-20726)-19597)+2740)+x^2 (y (y ((115656-64919 y) y-30207)+608)-108)\\
  \nonumber& &+2 x y^2 (7 y (1653 y-2510)+334)+4 y^2 (y (396 y-739)-27))+10 s^3 x^4 (y-1)^3 y^4 (1604 x^3 (y-1)^2+6 x^2 (y\\
  \nonumber& &-1) (99 y+560)+x ((7301-6388 y) y+1666)+y (4190 y-5929)+90))\},\\
  \nonumber\rho _{\langle \bar{q} Gq\rangle}^{1,\varSigma _c^*D}(s)&=&\langle g_s \bar{q}\sigma Gq\rangle \int_{x_{min}}^{x_{max}}  \,dx \int_{y_{min}}^{y_{max}}  \,dy \Bigg\{ \frac{-1}{589824 \pi^6 x^3 y^3}\{ (x-1) (m_c^5 (x (-y)+x+y)^2 (82 x^4 (y-1)^3+x^3 (y-1)^2 (27 y+\\
  \nonumber& &218)-x^2 (y-1) (y (783 y+184)-136)+x y (y (1466 y-2089)-145)-792 (y-2) y^2)-2 m_c^3 s x (y-1) y (x\\
  \nonumber& & (y-1)-y) (134 x^4 (y-1)^3-x^3 (y-1)^2 (33 y-322)-x^2 (y-1) (y (1239 y+284)-188)+x y (y (2458 y-3041\\
  \nonumber& &)-185)-264 y^2 (5 y-8))+3 m_c s^2 x^2 (y-1)^2 y^2 (62 x^4 (y-1)^3-x^3 (y-1)^2 (31 y-142)-x^2 (y-1) (y(565 y+ \\
  \nonumber& &128)-80)+x y (y (1150 y-1331)-75)+88 (10-7 y) y^2))\}+\frac{-1}{16384 \pi^6 x^2 y^2}\{ (x-1) (m_c^5 (x (-y)+x+y)^2\\
  \nonumber& & (63 x^2 (y-1)^2-x (90 y-61) (y-1)+y (27 y-49))-2 m_c^3 s x (y-1) y (x (y-1)-y) (83 x^2 (y-1)^2-x (130 y\\
  \nonumber& &-81) (y-1)+y (47 y-69))+m_c s^2 x^2 (y-1)^2 y^2 (103 x^2 (y-1)^2-x (170 y-101) (y-1)+y (67 y-89)))\}\Bigg\} ,\\
  \nonumber\rho _{\langle \bar{q} q\rangle^2}^{1,\varSigma _c^*D}(s)&=&\langle \bar{q} q\rangle^2 \int_{x_{min}}^{x_{max}}  \,dx \int_{y_{min}}^{y_{max}}  \,dy  \frac{1}{1536 \pi^4 x y}\{ (x-1) (y-1) (m_c^2 (-25 x (y-1)+25 y-42)+43 s x (y-1) y) (m_c^2 (x (\\
  \nonumber& &-y)+x+y)+s x (y-1) y) \},\\
  \nonumber\rho _{\langle \bar{q} q\rangle\langle \bar{q}G q\rangle}^{1,\varSigma _c^*D}(s)&=&\langle \bar{q} q\rangle \langle g_s^2 \bar{q}\sigma G q\rangle \int_{x_{min}}^{x_{max}}  \,dx \int_{y_{min}}^{y_{max}}  \,dy\Bigg\{  \frac{1}{1536 \pi^4}\{ (y-1) (m_c^2 (-25 x (y-1)+25 y-21)+34 s x (y-1) y)\}\\
  \nonumber& &+\frac{1}{36864 \pi^4 x y}\{ (y-1) (m_c^2 (-192 x^3 (y-1)^2+2 x^2 (y-1) (246 y-119)-x (y (388 y+233)+46)+4 y (22 y\\
  \nonumber& &+133))+s x (y-1) y (x (246 x (y-1)-371 y+246)+107 y))\}\Bigg\},\\
  \nonumber\rho _{\langle \bar{q}G q\rangle^2}^{1,\varSigma _c^*D}(s)&=&\langle g_s^2 \bar{q}\sigma G q\rangle ^2 \int_{x_{min}}^{x_{max}}  \,dx \int_{y_{min}}^{y_{max}}  \,dy  \Bigg\{\frac{1}{1769472 \pi^4}\{ (y-1) (9 x (y-1)-7 y-101)\}+\frac{1}{18432 \pi^4}\{ (y-1) (24 x (y-1)\\
  \nonumber& &-11 y)\}+\langle g_s^2 \bar{q}\sigma G q\rangle ^2 \int_{0}^{1}  \,dx \int_{0}^{1}  \,dy\Bigg\{ \frac{1}{1769472 \pi^4 x y}\{ m_c^2 (95 x^2 (y-1)^2-x (222 y-221) (y-1)+y (127 y\\
  \nonumber& &+665)+16)\}+\frac{-1}{147456 \pi^4 x y}\{ m_c^2 (54 x^2 (y-1)^2-x (73 y+46) (y-1)+19 (y-28) y)\}\Bigg\}\delta (s-\mathcal{F} (m_c^2))\},\\
  \rho _{\langle \bar{q} q\rangle^3}^{1,\varSigma _c^*D}(s)&=&\langle \bar{q} q\rangle ^3 \int_{x_{min}}^{x_{max}}  \,dx   \frac{1}{576 \pi^2}\{ m_c-12 m_c x \},
\end{eqnarray}
}

{\allowdisplaybreaks
\begin{eqnarray}
  \nonumber\rho _{pert}^{0,\varSigma _c^*D}(s)&=&\int_{x_{min}}^{x_{max}}  \,dx \int_{y_{min}}^{y_{max}}  \,dy \frac{1}{19660800 \pi^8 x^5 y^5}\{ (x-1)^4 (m_c^2 (x (y-1)-y)-s x (y-1) y)^4 (m_c^3 (x (y-1)-y) (32 x^2 (y-\\
  \nonumber& &1)^2-x (28 y+73) (y-1)-y (4 y+1))-2 m_c s x (y-1) y (41 x^2 (y-1)^2-x (34 y+149) (y-1)-y (7 y+8)))\},\\
  \nonumber\rho _{\langle \bar{q}q\rangle}^{0,\varSigma _c^*D}(s)&=&\langle \bar{q}q\rangle \int_{x_{min}}^{x_{max}}  \,dx \int_{y_{min}}^{y_{max}}  \,dy \frac{-1}{147456 \pi^6 x^3 y^3}\{ m_c (x-1)^2 (m_c^2 (x (-y)+x+y)+s x (y-1) y)^2 (3 m_c^4 (x (y-1)-y)\\
  \nonumber& & (41 x^2 (y-1)^2-x (82 y-31) (y-1)+y (41 y-31)+24)-2 m_c^2 s x (y-1) y (365 x^2 (y-1)^2-x (730 y\\
  \nonumber& &-359) (y-1)+y (365 y-359)+36)+s^2 x^2 (y-1)^2 y^2 (847 x (y-1)-847 y+865))\},\\
  \nonumber\rho _{\langle GG\rangle}^{0,\varSigma _c^*D}(s)&=&\langle g_s^2GG\rangle \int_{x_{min}}^{x_{max}}  \,dx \int_{y_{min}}^{y_{max}}  \,dy \frac{1}{283115520 \pi^8 x^5 y^5}\{ (x-1)^2 (-2 m_c^7 (x (-y)+x+y)^3 (384 x^6 (y-1)^4-28 x^5 (y-1)^3 \\
  \nonumber& &(43 y+6)+x^4 (y-1)^2 (y (857 y+800)-1488)+4 x^3 (y-1) (y (y (199 y+504)+368)-234)-x^2 y (y (y (595 y\\
  \nonumber& &+431)-1633)+532)-2 x y^2 (y (95 y+128)-238)-24 y^3 (2 y+3))-3 m_c^5 s x y (x (-y)+x+y)^2 (840 x^6 (y\\
  \nonumber& &-1)^5-4 x^5 (y-1)^4 (722 y+195)+x^4 (y-1)^3 (y (2165 y+3056)-4080)+2 x^3 (y-1)^2 (y (y (1134 y+3163)+\\
  \nonumber& &2398)-1230)-x^2 (y-1) y (y (y (1879 y+1800)-5257)+1148)-2 x (y-1) y^2 (y (193 y+279)-482)-60 y^3 \\
  \nonumber& &(2 y^2+y-3))+6 m_c^3 s^2 x^2 (y-1)^2 y^2 (416 x^7 (y-1)^5-4 x^6 (y-1)^4 (507 y+128)+8 x^5 (y-1)^3 (y (362 y+333)\\
  \nonumber& &-284)+2 x^4 (y-1)^2 (y (y (110 y+1081)+2806)-672)-x^3 (y-1) y (y (y (2860 y+5783)-284)-920)+x^2 \\
  \nonumber& &y^2 (y (y (1184 y+1467)-3154)+128)+2 x y^3 (y (54 y+71)-100)+32 y^4 (2 y+3))-m_c s^3 x^3 (y-1)^3 y^3 (744 x^6\\
  \nonumber& & (y-1)^4-4 x^5 (y-1)^3 (854 y+267)+x^4 (y-1)^2 (y (2923 y+5344)-4368)+2 x^3 (y-1) (y (y (1906 y+5469)\\
  \nonumber& &+4298)-1278)-x^2 y (y (y (3689 y+4276)-9239)+164)-2 x y^2 (y (127 y-59)+82)-60 y^3 (2 y+3)))\},\\
  \nonumber\rho _{\langle \bar{q} Gq\rangle}^{0,\varSigma _c^*D}(s)&=&\langle g_s \bar{q}\sigma Gq\rangle \int_{x_{min}}^{x_{max}}  \,dx \int_{y_{min}}^{y_{max}}  \,dy \Bigg\{ \frac{1}{1769472 \pi^6 x^3 y^3}\{ (x-1) (m_c^6 (x (-y)+x+y)^2 (38 x^4 (y-1)^3+x^3 (y-1)^2 (43 y\\
  \nonumber& &+112)-x^2 (y-1) (3 y (75 y+271)-182)+x (y (y (169 y+1506)+446)+108)-5 y (y (5 y+166)+324))- \\
  \nonumber& &9 m_c^4 s x (y-1) y (x (y-1)-y) (17 x^4 (y-1)^3+x^3 (y-1)^2 (11 y+63)-2 x^2 (y-1) (2 y (21 y+59)-35)+x (y (\\
  \nonumber& &y (67 y+360)+56)+24)-y (11 y (y+18)+360))+3 m_c^2 s^2 x^2 (y-1)^2 y^2 (68 x^4 (y-1)^3+x^3 (y-1)^2 (11 y+\\
  \nonumber& &274)-x^2 (y-1) (y (267 y+619)-242)+x (y (y (229 y+662)-114)+36)-y (y (41 y+358)+540))-s^3 x^3 (y \\
  \nonumber& &-1)^3 y^3 (89 x^3(y-1)^2+x^2 (y-1) (66 y+367)+x (278-y (204 y+179))+y (49 y+122)))\}+\frac{1}{49152 \pi^6 x^2 y^2}\{ (x\\
  \nonumber& &-1) (m_c^6 (x (-y)+x+y)^2 (123 x^2 (y-1)^2-2 x (123 y-50) (y-1)+y (123 y-100)+27)-6 m_c^4 s x (y-1) y (x \\
  \nonumber& & (y-1)-y) (122 x^2(y-1)^2-4 x (61 y-29) (y-1)+2 y (61 y-58)+9)+3 m_c^2 s^2 x^2 (y-1)^2 y^2 (405 x^2 (y-1)^2\\
  \nonumber& &-2 x (405 y-202) (y-1)+y (405 y-404)+9)-2 s^3 x^3 (y-1)^3 y^3 (303 x (y-1)-303 y+308))\}\Bigg\} ,\\
  \nonumber\rho _{\langle \bar{q} q\rangle^2}^{0,\varSigma _c^*D}(s)&=&\langle \bar{q} q\rangle^2 \int_{x_{min}}^{x_{max}}  \,dx \int_{y_{min}}^{y_{max}}  \,dy  \frac{1}{768 \pi^4 x^2 y^2}\{ m_c (x-1) (31 x (y-1)-19 y) (m_c^2 (x (-y)+x+y)+s x (y-1) y) \\
  \nonumber& &(m_c^2 (x (-y)+x+y)+2 s x (y-1) y) \},\\
  \nonumber\rho _{\langle \bar{q} q\rangle\langle \bar{q}G q\rangle}^{0,\varSigma _c^*D}(s)&=&\langle \bar{q} q\rangle \langle g_s^2 \bar{q}\sigma G q\rangle \int_{x_{min}}^{x_{max}}  \,dx \int_{y_{min}}^{y_{max}}  \,dy\Bigg\{  \frac{1}{1536 \pi^4 x y}\{ m_c (31 x (y-1)-19 y) (2 m_c^2 (x (-y)+x+y)+3 s x (y-1) y)\}\\
  \nonumber& &+\frac{1}{36864 \pi^4 x^2 y^2} 2 m_c s x (y-1) y (208 x^3 (y-1)^2-2 x^2 (y-1) (153 y-104)-x y (717 y+80)+799 y^2)-m_c^3 (x (y\\
  \nonumber& &-1)-y) (276 x^3 (y-1)^2-x^2 (y-1) (425 y-276)-x y (951 y+122)+1066 y^2)\}\Bigg\},\\
  \nonumber\rho _{\langle \bar{q}G q\rangle^2}^{0,\varSigma _c^*D}(s)&=&\langle g_s^2 \bar{q}\sigma G q\rangle ^2 \int_{x_{min}}^{x_{max}}  \,dx \int_{y_{min}}^{y_{max}}  \,dy  \Bigg\{\frac{1}{1769472 \pi^4 x y}\{ m_c (-112 x^2 (y-1)^2-x (599 y+188) (y-1)-2 y (33 y\\
  \nonumber& &+361))\}+\frac{1}{73728 \pi^4 x y}\{ (y-1) m_c (138 x^2 (y-1)^2-61 x y (y-1)-533 y^2)\}+\langle g_s^2 \bar{q}\sigma G q\rangle ^2 \int_{0}^{1}  \,dx \int_{0}^{1}  \,dy\Bigg\{  \\
  \nonumber& &\frac{1}{1769472 \pi^4 x^2 (y-1) y^2}\{ m_c^3 (x (y-1)-y) (8 x^2 (y-1)^2-x (77 y-30) (y-1)+4 y (5 y-7)) \}\\
  \nonumber& &+\frac{-1}{73728 \pi^4 x^2 (y-1) y^2}\{ m_c^3 (x (y-1)-y) (70 x^2 (y-1)^2-19 x y (y-1)-266 y^2)\}\Bigg\}\delta (s-\mathcal{F} (m_c^2))\},\\
  \rho _{\langle \bar{q} q\rangle^3}^{0,\varSigma _c^*D}(s)&=&\langle \bar{q} q\rangle ^3 \int_{x_{min}}^{x_{max}}  \,dx   \frac{-1}{192 \pi^2} 13 m_c^2+3 s (x-1) x \},
\end{eqnarray}
}

{\allowdisplaybreaks
\begin{eqnarray}
  \nonumber\rho _{pert}^{1,\varSigma _c^*D^*}(s)&=&\int_{x_{min}}^{x_{max}}  \,dx \int_{y_{min}}^{y_{max}}  \,dy \frac{1}{491520 \pi^8 x^4 y^4}\{ (x-1)^4 (y-1) (m_c^2 (x (-y)+x+y)+s x (y-1) y)^3 (m_c^4 (x (y-1)-y)\\
  \nonumber& & (3 x^2 (y-1)^2-6 x (y-2) (y-1)+3 (y-4) y-4)-m_c^2 s x (y-1) y (26 x^2 (y-1)^2-x (52 y-63) (y-1)\\
  \nonumber& &+y (26 y-63)-4)+s^2 x^2 (y-1)^2 y^2 (39 x (y-1)-39 y+67))\},\\
  \nonumber\rho _{\langle \bar{q}q\rangle}^{1,\varSigma _c^*D^*}(s)&=&\langle \bar{q}q\rangle \int_{x_{min}}^{x_{max}}  \,dx \int_{y_{min}}^{y_{max}}  \,dy \frac{1}{4608 \pi^6 x^3 y^3}\{ m_c (x-1)^2 (m_c^2 (x (-y)+x+y)+s x (y-1) y)^2 (m_c^2 (x (y-1)-y) (8 x^2 \\
  \nonumber& &(y-1)^2-x (13 y-8) (y-1)+y (5 y-14))-s x (y-1) y (20 x^2 (y-1)^2-x (37 y-20) (y-1)+y (17 y-26)))\},\\
  \nonumber\rho _{\langle GG\rangle}^{1,\varSigma _c^*D^*}(s)&=&\langle g_s^2GG\rangle \int_{x_{min}}^{x_{max}}  \,dx \int_{y_{min}}^{y_{max}}  \,dy \frac{-1}{70778880 \pi^8 x^4 y^4}\{ (x-1)^2 (y-1) (m_c^6 (x (-y)+x+y)^2 (360 x^6 (y-1)^4-5 x^5 (y\\
  \nonumber& &-1)^3 (241 y-432)+x^4 (y-1)^2 (17 y (10 y-321)+2760)+x^3 (y-1) (y (y (980 y+5021)-2981)+480)+\\
  \nonumber& &x^2 (y (y (y (910 y-7451)+1542)+1079)-480)+x y (y (3 (2104-525 y) y+1271)+192)+120 y^2 (3 (y-4) y\\
  \nonumber& &-4))-3 m_c^4 s x (y-1) y (x (y-1)-y) (560 x^6 (y-1)^4-4 x^5 (y-1)^3 (459 y-676)+x^4 (y-1)^2 (y (319 y-6819)\\
  \nonumber& &+3472)+x^3 (y-1) (y (y (779 y+6537)-4124)+1072)+x^2 (3 y (y (y (863 y-3619)+1283)+265)-256)+\\
  \nonumber& &x y (y ((8168-2971 y) y+859)+64)+16 y^2 (y (35 y-99)-16))+3 m_c^2 s^2 x^2 (y-1)^2 y^2 (504 x^6 (y-1)^4-x^5 (y\\
  \nonumber& &-1)^3 (1667 y-2176)-x^4 (y-1)^2 (y (508 y+5941)-2744)+x^3 (y-1) (y (y (1202 y+6853)-3731)+976)\\
  \nonumber& &+x^2 (y (y (y (3772 y-13543)+6204)+543)-96)+x y^2 ((8824-3807 y) y+543)+8 y^2 (y (63 y-146)\\
  \nonumber& &-12))+s^3 x^4 (y-1)^3 y^4 (122 x^3 (y-1)^2+3 x^2 (y-1) (1003 y+557)+2 x (y (188 y-1861)\\
  \nonumber& &+613)-3507 y^2+7704 y+323))\},\\
  \nonumber\rho _{\langle \bar{q} Gq\rangle}^{1,\varSigma _c^*D^*}(s)&=&\langle g_s \bar{q}\sigma Gq\rangle \int_{x_{min}}^{x_{max}}  \,dx \int_{y_{min}}^{y_{max}}  \,dy \Bigg\{ \frac{-1}{12288 \pi^6 x^2 y^2}\{ m_c (x-1)^2 (y-1) (17 x (y-1)-8 y+4) (m_c^2 (x (-y)+x+y)+\\
  \nonumber& &s x (y-1) y) (m_c^2 (x (-y)+x+y)+2 s x (y-1) y)\}+\frac{-1}{2048 \pi^6 x^2 y^2}\{ (x-1) (m_c^5 (x (-y)+x+y)^2 (8 x^2 (y-1)^2-x\\
  \nonumber& & (13 y-8) (y-1)+y (5 y-11))-6 m_c^3 s x (y-1) y (x (y-1)-y) (4 x^2 (y-1)^2-x (7 y-4) (y-1)+y (3 y-5))\\
  \nonumber& &+m_c s^2 x^2 (y-1)^2 y^2 (16 x^2 (y-1)^2-x (29 y-16) (y-1)+y (13 y-19)))\}\Bigg\} ,\\
  \nonumber\rho _{\langle \bar{q} q\rangle^2}^{1,\varSigma _c^*D^*}(s)&=&\langle \bar{q} q\rangle^2 \int_{x_{min}}^{x_{max}}  \,dx \int_{y_{min}}^{y_{max}}  \,dy  \frac{1}{96 \pi^4 x y}\{ (x-1) (y-1) (m_c^2 (-2 x (y-1)+2 y-5)+4 s x (y-1) y) (m_c^2 (x (-y)\\
  \nonumber& &+x+y)+s x (y-1) y) \},\\
  \nonumber\rho _{\langle \bar{q} q\rangle\langle \bar{q}G q\rangle}^{1,\varSigma _c^*D^*}(s)&=&\langle \bar{q} q\rangle \langle g_s^2 \bar{q}\sigma G q\rangle \int_{x_{min}}^{x_{max}}  \,dx \int_{y_{min}}^{y_{max}}  \,dy\Bigg\{  \frac{1}{192 \pi^4}\{ (y-1) (m_c^2 (-4 x (y-1)+4 y-5)+6 s x (y-1) y)\}+\frac{1}{1536 \pi^4 x y}\\
  \nonumber& & \{(y-1) (3 s (x-1) x (y-1) y (x (y-1)+2 y)-m_c^2 (2 x^3 (y-1)^2+2 x^2 (y-1)+x (5-6 y) y+4 y^2))\}\Bigg\},\\
  \nonumber\rho _{\langle \bar{q}G q\rangle^2}^{1,\varSigma _c^*D^*}(s)&=&\langle g_s^2 \bar{q}\sigma G q\rangle ^2 \int_{x_{min}}^{x_{max}}  \,dx \int_{y_{min}}^{y_{max}}  \,dy  \Bigg\{\frac{1}{221184 \pi^4}\{ (y-1) (30 x (y-1)-43 y+33)\}+\frac{1}{3072 \pi^4}\{ (y-1) (x (y-1)+2 y)\}\\
  \nonumber& &+\langle g_s^2 \bar{q}\sigma G q\rangle ^2 \int_{0}^{1}  \,dx \int_{0}^{1}  \,dy\Bigg\{ \frac{-1}{442368 \pi^4 x y}\{ m_c^2 (8 x (y-1)-11 y+8) (4 x (y-1)-4 y+3)\}\\
  \nonumber& &+\frac{-1}{6144 \pi^4 x y}\{ m_c^2 (x (y-1)-y) (x (y-1)+2 y)\}\Bigg\}\delta (s-\mathcal{F} (m_c^2))\},\\
 \rho _{\langle \bar{q} q\rangle^3}^{1,\varSigma _c^*D^*}(s)&=&\langle \bar{q} q\rangle ^3 \int_{x_{min}}^{x_{max}}  \,dx   \frac{-1}{24 \pi^2}\{ m_c x \},
\end{eqnarray}
}

{\allowdisplaybreaks
\begin{eqnarray}
  \nonumber\rho _{pert}^{1,\varSigma _c^*D^*}(s)&=&\int_{x_{min}}^{x_{max}}  \,dx \int_{y_{min}}^{y_{max}}  \,dy \frac{1}{491520 \pi^8 x^4 y^4} (x-1)^4 (y-1) (m_c^2 (x (-y)+x+y)+s x (y-1) y)^3 (m_c^4 (x (y-1)-y)\\
  \nonumber& & (3 x^2 (y-1)^2-6 x (y-2) (y-1)+3 (y-4) y-4)-m_c^2 s x (y-1) y (26 x^2 (y-1)^2-x (52 y-63) (y-1)\\
  \nonumber& &+y (26 y-63)-4)+s^2 x^2 (y-1)^2 y^2 (39 x (y-1)-39 y+67)),\\
  \nonumber\rho _{\langle \bar{q}q\rangle}^{1,\varSigma _c^*D^*}(s)&=&\langle \bar{q}q\rangle \int_{x_{min}}^{x_{max}}  \,dx \int_{y_{min}}^{y_{max}}  \,dy \frac{1}{4608 \pi^6 x^3 y^3} m_c (x-1)^2 (m_c^2 (x (-y)+x+y)+s x (y-1) y)^2 (m_c^2 (x (y-1)-y) (8 x^2 \\
  \nonumber& &(y-1)^2-x (13 y-8) (y-1)+y (5 y-14))-s x (y-1) y (20 x^2 (y-1)^2-x (37 y-20) (y-1)+y (17 y-26))),\\
  \nonumber\rho _{\langle GG\rangle}^{1,\varSigma _c^*D^*}(s)&=&\langle g_s^2GG\rangle \int_{x_{min}}^{x_{max}}  \,dx \int_{y_{min}}^{y_{max}}  \,dy \frac{-1}{70778880 \pi^8 x^4 y^4} (x-1)^2 (y-1) (m_c^6 (x (-y)+x+y)^2 (360 x^6 (y-1)^4-5 x^5 (y\\
  \nonumber& &-1)^3 (241 y-432)+x^4 (y-1)^2 (17 y (10 y-321)+2760)+x^3 (y-1) (y (y (980 y+5021)-2981)+480)+\\
  \nonumber& &x^2 (y (y (y (910 y-7451)+1542)+1079)-480)+x y (y (3 (2104-525 y) y+1271)+192)+120 y^2 (3 (y-4) y\\
  \nonumber& &-4))-3 m_c^4 s x (y-1) y (x (y-1)-y) (560 x^6 (y-1)^4-4 x^5 (y-1)^3 (459 y-676)+x^4 (y-1)^2 (y (319 y-6819)\\
  \nonumber& &+3472)+x^3 (y-1) (y (y (779 y+6537)-4124)+1072)+x^2 (3 y (y (y (863 y-3619)+1283)+265)-256)+\\
  \nonumber& &x y (y ((8168-2971 y) y+859)+64)+16 y^2 (y (35 y-99)-16))+3 m_c^2 s^2 x^2 (y-1)^2 y^2 (504 x^6 (y-1)^4-x^5 (y\\
  \nonumber& &-1)^3 (1667 y-2176)-x^4 (y-1)^2 (y (508 y+5941)-2744)+x^3 (y-1) (y (y (1202 y+6853)-3731)+976)\\
  \nonumber& &+x^2 (y (y (y (3772 y-13543)+6204)+543)-96)+x y^2 ((8824-3807 y) y+543)+8 y^2 (y (63 y-146)\\
  \nonumber& &-12))+s^3 x^4 (y-1)^3 y^4 (122 x^3 (y-1)^2+3 x^2 (y-1) (1003 y+557)+2 x (y (188 y-1861)\\
  \nonumber& &+613)-3507 y^2+7704 y+323)),\\
  \nonumber\rho _{\langle \bar{q} Gq\rangle}^{1,\varSigma _c^*D^*}(s)&=&\langle g_s \bar{q}\sigma Gq\rangle \int_{x_{min}}^{x_{max}}  \,dx \int_{y_{min}}^{y_{max}}  \,dy \Bigg\{ \frac{-1}{12288 \pi^6 x^2 y^2} m_c (x-1)^2 (y-1) (17 x (y-1)-8 y+4) (m_c^2 (x (-y)+x+y)+\\
  \nonumber& &s x (y-1) y) (m_c^2 (x (-y)+x+y)+2 s x (y-1) y)+\frac{-1}{2048 \pi^6 x^2 y^2} (x-1) (m_c^5 (x (-y)+x+y)^2 (8 x^2 (y-1)^2-x\\
  \nonumber& & (13 y-8) (y-1)+y (5 y-11))-6 m_c^3 s x (y-1) y (x (y-1)-y) (4 x^2 (y-1)^2-x (7 y-4) (y-1)+y (3 y-5))\\
  \nonumber& &+m_c s^2 x^2 (y-1)^2 y^2 (16 x^2 (y-1)^2-x (29 y-16) (y-1)+y (13 y-19)))\Bigg\} ,\\
  \nonumber\rho _{\langle \bar{q} q\rangle^2}^{1,\varSigma _c^*D^*}(s)&=&\langle \bar{q} q\rangle^2 \int_{x_{min}}^{x_{max}}  \,dx \int_{y_{min}}^{y_{max}}  \,dy  \frac{1}{96 \pi^4 x y} (x-1) (y-1) (m_c^2 (-2 x (y-1)+2 y-5)+4 s x (y-1) y) (m_c^2 (x (-y)\\
  \nonumber& &+x+y)+s x (y-1) y) ,\\
  \nonumber\rho _{\langle \bar{q} q\rangle\langle \bar{q}G q\rangle}^{1,\varSigma _c^*D^*}(s)&=&\langle \bar{q} q\rangle \langle g_s^2 \bar{q}\sigma G q\rangle \int_{x_{min}}^{x_{max}}  \,dx \int_{y_{min}}^{y_{max}}  \,dy\Bigg\{  \frac{1}{192 \pi^4} (y-1) (m_c^2 (-4 x (y-1)+4 y-5)+6 s x (y-1) y)+\frac{1}{1536 \pi^4 x y}\\
  \nonumber& & (y-1) (3 s (x-1) x (y-1) y (x (y-1)+2 y)-m_c^2 (2 x^3 (y-1)^2+2 x^2 (y-1)+x (5-6 y) y+4 y^2))\Bigg\},\\
  \nonumber\rho _{\langle \bar{q}G q\rangle^2}^{1,\varSigma _c^*D^*}(s)&=&\langle g_s^2 \bar{q}\sigma G q\rangle ^2 \int_{x_{min}}^{x_{max}}  \,dx \int_{y_{min}}^{y_{max}}  \,dy  \Bigg\{\frac{1}{221184 \pi^4} (y-1) (30 x (y-1)-43 y+33)+\frac{1}{3072 \pi^4} (y-1) (x (y-1)+2 y)\\
  \nonumber& &+\langle g_s^2 \bar{q}\sigma G q\rangle ^2 \int_{0}^{1}  \,dx \int_{0}^{1}  \,dy\Bigg\{ \frac{-1}{442368 \pi^4 x y} m_c^2 (8 x (y-1)-11 y+8) (4 x (y-1)-4 y+3)\\
  \nonumber& &+\frac{-1}{6144 \pi^4 x y} m_c^2 (x (y-1)-y) (x (y-1)+2 y)\Bigg\}\delta (s-\mathcal{F} (m_c^2)),\\
\rho _{\langle \bar{q} q\rangle^3}^{1,\varSigma _c^*D^*}(s)&=&\langle \bar{q} q\rangle ^3 \int_{x_{min}}^{x_{max}}  \,dx   \frac{-1}{24 \pi^2} m_c x ,
\end{eqnarray}
}

{\allowdisplaybreaks
\begin{eqnarray}
  \nonumber\rho _{pert}^{0,\varSigma _c^*D^*}(s)&=&\int_{x_{min}}^{x_{max}}  \,dx \int_{y_{min}}^{y_{max}}  \,dy \frac{-1}{819200 \pi^8 x^4 y^5} \Bigg\{m_c (x-1)^4 (y-1) (x (y-1)-y-4) (m_c^2 (x (-y)+x+y)+s x (y-1) y)^4\\
  \nonumber& & (m_c^2 (x (-y)+x+y)+6 s x (y-1) y)\Bigg\},\\
  \nonumber\rho _{\langle \bar{q}q\rangle}^{0,\varSigma _c^*D^*}(s)&=&\langle \bar{q}q\rangle \int_{x_{min}}^{x_{max}}  \,dx \int_{y_{min}}^{y_{max}}  \,dy \frac{-1}{36864 \pi^6 x^3 y^3} \Bigg\{(x-1)^2 (m_c^2 (x (-y)+x+y)+s x (y-1) y)^2 (m_c^4 (x (y-1)-y) (39 x^2\\
  \nonumber& & (y-1)^2-26 x (3 y-1) (y-1)+13 y (3 y-2)+32)-2 m_c^2 s x (y-1) y (123 x^2 (y-1)^2-2 x (123 y-65)\\
  \nonumber& & (y-1)+y (123 y-130)+16)+3 s^2 x^2 (y-1)^2 y^2 (101 x (y-1)-101 y+110))\}\Bigg\},\\
  \nonumber\rho _{\langle GG\rangle}^{0,\varSigma _c^*D^*}(s)&=&\langle g_s^2GG\rangle \int_{x_{min}}^{x_{max}}  \,dx \int_{y_{min}}^{y_{max}}  \,dy \frac{1}{141557760 \pi^8 x^4 y^5} \Bigg\{(x-1)^2 (m_c^7 (-(x (-y)+x+y)^3) (288 x^5 (y-1)^4-2 x^4 (y-1)^3 \\
  \nonumber& &(355 y+288)+x^3 (y-1)^2 (y (2059 y+1644)-2016)-2 x^2 (y-1) (y (17 y (83 y-19)-889)+576)+x y (13 y \\
  \nonumber& &(y (99 y+70)-137)-576)-2 y^2 (y (51 y+163)-294))-3 m_c^5 s x (y-1) y (x (-y)+x+y)^2 (480 x^5 (y-1)^4-\\
  \nonumber& &2 x^4 (y-1)^3 (679 y+480)+x^3 (y-1)^2 (y (3499 y+3036)-3360)-2 x^2 (y-1) (y (y (2047 y-719)-1813)+\\
  \nonumber& &960)+x y (y (y (1623 y+1486)-2501)-768)-10 y^2 (3 y+13) (5 y-6))+3 m_c^3 s^2 x^2 (y-1)^2 y^2 (576 x^6 (y-1)^5\\
  \nonumber& &-2 x^5 (y-1)^4 (1195 y+576)+3 x^4 (y-1)^3 (y (2219 y+1796)-1344)-x^3 (y-1)^2 (y (y (10113 y+2006)-\\
  \nonumber& &9506)+2304)+x^2 (y-1) y (y (y (7037 y-360)-8695)+1728)-x y^2 (y (3 y (623 y+698)-4135)+12)+2 y^3\\
  \nonumber& & (y (51 y+163)-294))-m_c s^3 x^3 (y-1)^3 y^3 (576 x^5 (y-1)^4-2 x^4 (y-1)^3 (1039 y+576)+x^3 (y-1)^2 (y (6091 y\\
  \nonumber& &+5244)-4032)-2 x^2 (y-1) (y (y (3175 y-1511)-3661)+1152)+x y^2 (y (1719 y+2062)\\
  \nonumber& &-3941)+2 y^2 (y (21 y+53)+6)))\Bigg\}, \\
  \nonumber\rho _{\langle \bar{q} Gq\rangle}^{0,\varSigma _c^*D^*}(s)&=&\langle g_s \bar{q}\sigma Gq\rangle \int_{x_{min}}^{x_{max}}  \,dx \int_{y_{min}}^{y_{max}}  \,dy \Bigg\{ \frac{-1}{221184 \pi^6 x^2 y^3} \Bigg\{(x-1)^2 (y-1) (s x (y-1) y (44 x (y-1)+97 y+47)-m_c^2 (x (y\\
  \nonumber& &-1)-y) (17 x (y-1)+43 y+20)) (m_c^2 (x (-y)+x+y)+s x (y-1) y)^2\Bigg\}+\frac{1}{12288 \pi^6 x^2 y^2}\Bigg\{(x-1) (m_c^6 (x (-y)+\\
  \nonumber& &x+y)^2 (39 x^2 (y-1)^2-x (78 y-29) (y-1)+y (39 y-29)+12)-3 m_c^4 s x (y-1) y (x (y-1)-y) (81 x^2 (y-1)^2\\
  \nonumber& &-x (162 y-79) (y-1)+y (81 y-79)+8)+3 m_c^2 s^2 x^2 (y-1)^2 y^2 (139 x^2 (y-1)^2-x (278 y-145) (y-1)+y \\
  \nonumber& &(139 y-145)+4)-s^3 x^3 (y-1)^3 y^3 (213 x (y-1)-213 y+227))\Bigg\}\Bigg\} ,\\
  \nonumber\rho _{\langle \bar{q} q\rangle^2}^{0,\varSigma _c^*D^*}(s)&=&\langle \bar{q} q\rangle^2 \int_{x_{min}}^{x_{max}}  \,dx \int_{y_{min}}^{y_{max}}  \,dy  \frac{1}{192 \pi^4 x^2 y^2} \Bigg\{(x-1) (m_c^5 (7 x (y-1)-10 y) (x (-y)+x+y)^2-28 m_c^3 s x (y-1) y (x (\\
  \nonumber& &-y)+x+y)^2+3 m_c s^2 x^2 (y-1)^2 y^2 (7 x (y-1)-6 y))\Bigg\} ,\\
  \nonumber\rho _{\langle \bar{q} q\rangle\langle \bar{q}G q\rangle}^{0,\varSigma _c^*D^*}(s)&=&\langle \bar{q} q\rangle \langle g_s^2 \bar{q}\sigma G q\rangle \int_{x_{min}}^{x_{max}}  \,dx \int_{y_{min}}^{y_{max}}  \,dy\Bigg\{  \frac{1}{192 \pi^4 x y} \Bigg\{m_c (x (y-1)-y) (m_c^2 (10 y-7 x (y-1))+14 s x (y-1) y)\Bigg\}\\
  \nonumber& &+\frac{1}{9216 \pi^4 x y^2}\Bigg\{m_c s x (y-1) y (96 x^2 (y-1)^2-x (47 y-96) (y-1)+(32-67 y) y)-m_c^3 (x (y-1)-y) (48 x^2 \\
  \nonumber& &(y-1)^2-x (31 y-48) (y-1)+(8-35 y) y)\Bigg\}\Bigg\},\\
  \nonumber\rho _{\langle \bar{q}G q\rangle^2}^{0,\varSigma _c^*D^*}(s)&=&\langle g_s^2 \bar{q}\sigma G q\rangle ^2 \int_{x_{min}}^{x_{max}}  \,dx \int_{y_{min}}^{y_{max}}  \,dy  \Bigg\{\frac{-1}{442368 \pi^4 x y} \{m_c (8 x^2 (y-1)^2+18 x y (y-1)+6 y^2+y)\}+\frac{1}{4608 \pi^4 y} \{m_c (y\\
  \nonumber& &-1) (6 x (y-1)+y)\}+\langle g_s^2 \bar{q}\sigma G q\rangle ^2 \int_{0}^{1}  \,dx \int_{0}^{1}  \,dy\Bigg\{ \frac{1}{73728 \pi^4 x y} \{m_c^3 (x (-y)+x+y)\}+\frac{-1}{1536 \pi^4 x y^2} \{m_c^3 (x (y\\
  \nonumber& &-1)-y) (2 x (y-1)+y)\}\Bigg\}\delta (s-\mathcal{F} (m_c^2)),\\
\rho _{\langle \bar{q} q\rangle^3}^{0,\varSigma _c^*D^*}(s)&=&\langle \bar{q} q\rangle ^3 \int_{x_{min}}^{x_{max}}  \,dx   \frac{-1}{144 \pi^2} \{19 m_c^2+5 s (x-1) x\} ,
\end{eqnarray}
}

\end{document}